\documentclass[titlepage,11pt]{article}

\usepackage{preamble}
\usepackage{array}

\begin{document}

\begin{titlepage}

\title{NLP-Informed Dynamic Cognitive Diagnosis Modelling}


\author{
Yawen Ma $^{1}$,
Sahoko Ishida $^{2}$,
Kate Cain $^{3}$ ,
Gabriel Wallin $^{4*}$
}

\affil{
$^{1}$School of Mathematical Sciences, Lancaster University\\
$^{2}$Department of Computer Science, University of Oxford\\
$^{3}$Department of Psychology, Lancaster University\\
$^{4}$School of Mathematical Sciences, Lancaster University
}


\linespacing{1}
\contact{$^{*}$Correspondence should be sent to Gabriel Wallin\\
E-Mail: g.wallin@lancaster.ac.uk}

\end{titlepage}

\setcounter{page}{2}
\vspace*{2\baselineskip}

\RepeatTitle{NLP-Informed Dynamic Cognitive Diagnosis Modelling}\vskip3pt

\linespacing{1.5}
\abstracthead
\begin{abstract}
Digital learning platforms are increasingly used to support reading development while generating rich log files and item-level textual content. Using these data, this study proposes a dynamic cognitive diagnostic modelling (CDM) framework that incorporates text-derived semantic information to inform the estimation of the $Q$-matrix. We construct item-level semantic representations of question text and response options, and use these representations to define an informative prior on the $Q$-matrix. This approach treats text-derived signals as proxies for item complexity and cognitive demands, guiding the item-skill mapping in a data-driven manner. The proposed framework jointly estimates latent skill mastery profiles, item parameters, and transition dynamics over time within a Bayesian framework. We apply the model to data from Boost Reading, a digital reading supplement, focusing on students’ vocabulary and comprehension skill development. We compare the proposed framework with a baseline model without any text information and show that the text-derived prior can improve Q-matrix recovery, particularly in settings where response data alone provide limited identification, as well as other model parameters for varying scenarios. This study provides a novel integration of natural language processing and dynamic CDMs, offering a data-driven approach to modelling skill acquisition and item–skill relationships in digital learning environments.

\end{abstract}\vspace{\fill}\pagebreak


\section{Introduction}

Digital learning environments are increasingly used to support the development of early reading skills, particularly in settings where teachers and schools seek more adaptive and individualised forms of instruction. In addition to providing learners with repeated practice and feedback, such platforms generate detailed records of student interactions, including response accuracy, timing information, reattempt patterns, and progression through activities. These data make it possible to study not only whether learners succeed or fail, but also how underlying skills develop across repeated interactions over time. A natural framework for analysing such data is provided by cognitive diagnostic models (CDMs) \citep{haertel1984application, junker2001cognitive}. CDMs aim to classify learners according to mastery or non-mastery of a set of latent attributes and to relate those latent mastery profiles to observed item responses. This is highly appealing in education because the goal is often not merely to produce an overall score, but to obtain interpretable information about specific skills that could guide teaching, intervention, and feedback. When the data are collected longitudinally, CDMs can also be extended to describe how mastery changes over time, which make them especially relevant for digital learning settings.

A central challenge in CDMs is the specification of the $Q$-matrix, the binary design matrix that indicates which attributes are required by each item. The $Q$-matrix is fundamental because it determines the substantive interpretation of the latent attributes and directly affects classification and parameter estimation. It is therefore well established that errors in the $Q$-matrix can lead to distorted inferences about both items and learners \citep{rupp2008effects, chen2015statistical}. Although many applications rely on expert-specified $Q$-matrices, such information is not always available in operational learning systems, and, even when it is, uncertainty may remain about whether particular items measure one skill or several. This difficulty is also present in dynamic settings, where instability in the estimated item--attribute structure can affect the interpretation of longitudinal learning trajectories.

\cite{ma2026dynamic} proposed a Bayesian dynamic cognitive diagnosis framework for digital learning data that jointly estimates time-varying latent attribute profiles, item parameters, covariate effects, and the unknown $Q$-matrix within a single model. The framework showed that it is possible to recover meaningful latent skill structures from log-file data without assuming that the item--attribute mapping is known in advance. At the same time, in settings where the response data are only moderately informative, the model may face uncertainty about the $Q$-matrix, especially when distinguishing between simpler and more complex item structures. The present paper is motivated directly by that problem.

The key idea of this current study is to bring in an additional source of information: the text of the items themselves. More specifically, we ask whether natural language processing (NLP) can provide useful prior information about the complexity of an item, and thereby about the plausible form of the $Q$-matrix. Importantly, our aim is not to use text to determine which specific attribute an item measures. Text-derived information is instead treated as evidence about whether an item is more likely to require relatively few attributes or multiple attributes. In this sense, the NLP component informs the prior distribution of the $Q$-matrix structure without replacing the response-based evidence that remains central to the diagnostic model. Recent advances in NLP make this type of extension increasingly feasible. Transformer-based language models have enabled semantic representations of text that capture contextual meaning beyond simple word overlap \citep{vaswani2017attention, devlin2019bert}. Sentence-level embedding methods such as Sentence-BERT (Sentence-Bidirectional Encoder Representations from Transformers (SBERT)) are particularly useful for similarity-based applications because they provide vector representations of texts while preserving semantic relationships \citep{reimers2019sentencebert}. More broadly, NLP and machine learning methods are now playing an important role in educational measurement and computational psychometrics, including applications in automated scoring, item generation, tutoring systems, and the analysis of assessment content \citep{von2018automated, gierl2012role, du2017learning, flor2022text, hommel2022transformer, von2021computational, martinkova2023computational}. These developments suggest that item wording may contain information that is relevant for psychometric modelling, even when that information is not strong enough to identify exact item--attribute mappings on its own.

Building on these ideas, we extend the Bayesian dynamic CDM framework of \cite{ma2026dynamic} by incorporating an NLP-derived item-level signal into the prior on the $Q$-matrix. The resulting model retains the original joint structure for estimating latent mastery profiles, slipping and guessing parameters, and transition effects, but augments the prior information used when learning the item--attribute structure. This allows us to examine whether text-derived information can stabilise the $Q$-matrix estimation in situations where response data alone leave ambiguity about whether an item is relatively simple or cognitively more demanding. In doing so, the paper contributes both to dynamic diagnostic modelling and to the broader goal of integrating AI-based tools into interpretable statistical models for educational data.

Using the proposed framework, we analyse data from Boost Reading (formerly Amplify Reading), which is an educational game-based reading supplementary developed by Amplify that has been widely implemented in the United States since 2018 (Amplify; \url{https://amplify.com/programs/boost-reading/}). As of 2024, it has been adopted by over a thousand school districts and serves more than one million students. It provides multiple games targeting various reading skills, such as phonological awareness, decoding, vocabulary and language comprehension, which are the key components of Simple View of Reading \citep{gough1986}. Evidence on the effectiveness of Boost Reading has been reported in \cite{newton2019examining} as well as in internal reports\footnote{See \url{https://amplify.com/research-and-case-studies/boost-reading-research/}.} (e.g., \citet{zoski2023}). Our previous work \citep{ma2026dynamic}, which utilized log files from Boost Reading, demonstrated the effectiveness of the dynamic CDM framework and showed strong recovery performance in simulation studies. Building on this work, the present study also uses data from Boost Reading as well, but focuses on incorporating text-based prior information to improve the estimation of the $Q$-matrix. 

The remainder of the paper is organised as follows. In Section~2, we describe the digital learning environment and the data structure that motivate the analysis. In Section~3, we present the baseline dynamic CDM framework and introduce the proposed NLP-informed prior for the $Q$-matrix. Section~4 reports the empirical application to Boost Reading data. Section~5 presents a simulation study designed to evaluate the extent to which text-informed priors improve recovery under varying levels of sample size, test length, and sparsity. Section~6 concludes with a discussion of implications, limitations, and directions for future work at the intersection of diagnostic modelling and NLP.

\section{Methodology}

\subsection{Dynamic Cognitive Diagnosis Model}

Let $Y_{ijt} \in \{0,1\}$ denote the binary response of learner $i$ to
item $j$ at time point $t$, with $Y_{ijt}=1$ indicating a correct response.
We model these responses using a dynamic CDM comprising a measurement model
that links observed responses to latent skill states and a structural model
that governs how those skill states evolve over time, following the framework
of \cite{ma2026dynamic}.

\paragraph{Measurement model.}
The latent state of learner $i$ at time $t$ is represented by an attribute
profile $\boldsymbol{\alpha}_{it} = (\alpha_{i1t},\ldots,\alpha_{iKt})^\top$,
where $\alpha_{ikt} \in \{0,1\}$ denotes mastery ($1$) or non-mastery ($0$)
of attribute $k$. The relationship between items and attributes is encoded in
the $J \times K$ binary $Q$-matrix, whose entry $q_{jk} = 1$ if item $j$
requires attribute $k$ and $q_{jk} = 0$ otherwise; the $Q$-matrix is treated
as unknown and estimated jointly within the model.
For each learner--item--time combination we define the ideal response
indicator
\begin{equation}
  \eta_{ijt} = \prod_{k=1}^{K}\alpha_{ikt}^{q_{jk}},
\end{equation}
which equals one if and only if learner $i$ has mastered every attribute
required by item $j$ at time $t$, and zero otherwise.
Under the Deterministic Inputs, Noisy And gate (DINA) model, the
probability of a correct response is
\begin{equation}
  P(Y_{ijt}=1 \mid \eta_{ijt}, g_j, s_j)
  = (1-s_j)^{\eta_{ijt}}\, g_j^{\smash{1-\eta_{ijt}}},
\end{equation}
where $s_j \in (0,1)$ is the slipping parameter (probability of an incorrect
response despite full attribute mastery) and $g_j \in (0,1)$ is the guessing
parameter (probability of a correct response in the absence of full mastery).

\paragraph{Structural model.}
Attribute mastery is assumed to follow a first-order Markov process.
Let $\mathbf{Z}_{it} = (Z_{it,1},\ldots,Z_{it,C})^\top$ denote the covariate
vector for learner $i$ at time $t$. The initial mastery probability at
$t = 1$ is specified as
\begin{equation}
  \label{eq:beta_initial}
  \operatorname{logit}\bigl(P(\alpha_{ik1}=1 \mid \mathbf{Z}_{i0})\bigr)
  = \beta_{0k} + \sum_{c=1}^{C}\beta_{kc}\,Z_{i0,c},
\end{equation}
where $\beta_{0k}$ is an intercept and $\beta_{kc}$ captures the effect of
covariate $c$ on initial mastery of attribute $k$.
Attribute transitions are parameterised as a logistic regression for the probability of gaining mastery between $t-1$ and $t$:
\begin{align*}
 \operatorname{logit}\bigl(P(\alpha_{ikt}=&1 \mid \alpha_{ik,t-1}=0, \mathbf{Z}_{i,t-1})\bigr)
  = \gamma_{01,k,0} + \sum_{c=1}^{C}\gamma_{01,k,c}\,Z_{i,t-1,c}, \\
  \operatorname{logit}\bigl(P(\alpha_{ikt}=&0 \mid \alpha_{ik,t-1}=1, \mathbf{Z}_{i,t-1})\bigr)
  = \gamma_{10,k,0} + \sum_{c=1}^{C}\gamma_{10,k,c}\,Z_{i,t-1,c}.
\end{align*}

The prior on the loss-of-mastery transition parameter $\gamma_{10}$ is specified to place most mass on low transition probabilities away from mastery, reflecting the expectation that consolidated early reading skills are rarely lost. Under this specification, apparent errors among proficient learners are attributed primarily to slipping rather than genuine mastery loss, though the model does not impose a hard absorbing-state constraint.

\subsection{Text-Derived Item Signal}

Each item in the assessment consists of a question stem, a correct answer option, and a set of distractors. Beyond indicating whether a response is correct, this text contains information about how cognitively demanding the item is likely to be. An item whose correct option is semantically well-separated from the distractors presents the learner with a clear discriminative signal, suggesting that the item targets a relatively focused skill. An item where distractors are semantically close to the correct option demands finer distinctions, which is consistent with a more complex or multi-faceted cognitive requirement. We formalise this intuition 
as an item-level semantic discriminability score, which is then used to inform the prior on the Q-matrix.

To construct this score, we represent item texts as dense vector embeddings using SBERT \citep[][]{reimers2019sentencebert}, an extension of the BERT transformer architecture \citep{devlin2019bert} specifically optimised to produce sentence-level representations in which semantic similarity corresponds to geometric proximity. Formally, SBERT maps each text segment to a vector $\mathbf{e} \in \mathbb{R}^d$. If two text segments are represented by embeddings $\mathbf{u}$ and $\mathbf{v}$, their similarity is measured by cosine similarity,
\begin{equation}
  \mathrm{sim}(\mathbf{u}, \mathbf{v})
  = \frac{\mathbf{u}^\top \mathbf{v}}{\|\mathbf{u}\|\,\|\mathbf{v}\|},
\end{equation}
which takes values in $[-1,1]$, with values close to one indicating high semantic similarity.

For item $j$, let $\mathbf{e}^{(q)}_j$, $\mathbf{e}^{(c)}_j$, and $\mathbf{e}^{(d_m)}_j$ denote the embeddings of the question stem, the correct option, and the $m$-th of $M_j$ distractors, respectively. We compute the similarity between the question stem and the correct option,
\begin{equation}
  S_j^{+} = \mathrm{sim}\!\left(\mathbf{e}^{(q)}_j, \mathbf{e}^{(c)}_j\right),
\end{equation}
and the mean similarity between the question stem and the distractors,
\begin{equation}
  S_j^{-} = \frac{1}{M_j}\sum_{m=1}^{M_j}
    \mathrm{sim}\!\left(\mathbf{e}^{(q)}_j, \mathbf{e}^{(d_m)}_j\right).
\end{equation}
The item-level text signal is then defined as the difference
\begin{equation}
  \tau_j = S_j^{+} - S_j^{-}.
\end{equation}
A positive value of $\tau_j$ indicates that the correct option is more semantically aligned with the question than are the distractors, representing high semantic discriminability. A value near zero indicates that the correct and incorrect options are equally close to the question text in the embedding space, reflecting that the semantic contrast is limited. Before entering the model, $\tau_j$ is standardised to have mean 0 and variance 1, reducing sensitivity to construction-induced scale differences across items with different numbers of distractors or different text structures. 

The main assumption linking this quantity to the Q-matrix structure is that items with higher semantic discriminability are more likely to target a focused set of attributes, and hence are more likely to have sparse Q-matrix rows, while items with lower semantic discriminability are more likely to require multiple attributes. 
This assumption is motivated by feature-based accounts of semantic memory, in which the similarity between attributes depends on the overlap and correlation among their features, whereas discriminability depends on the availability of distinctive features \citep{kumar2021, smith1974, tversky1977}. Under this view, items with lower semantic discriminability are likely to share more overlapping semantic features with distractors and therefore require more information to be uniquely identified.

\subsection{Text-Informed Prior for the $Q$-matrix}

Each entry $q_{jk}$ of the $Q$-matrix is assigned a Bernoulli prior whose success probability is informed by the item-level text signal $\tau_j$ introduced in the previous subsection:
\begin{equation}
  q_{jk} \sim \mathrm{Bernoulli}(\pi_{jk}),
  \qquad
  \mathrm{logit}(\pi_{jk}) = \mathrm{logit}(\theta) - \lambda \tau_j.
  \label{eq:lambda_T}
\end{equation}
The parameter $\theta \in (0,1)$ controls the overall sparsity of the $Q$-matrix, and $\lambda \geq 0$ governs how strongly $\tau_j$ shifts the prior inclusion probability. Because $\tau_j$ is standardised to have mean 0 and variance 1 and enters with a negative sign, items with higher-than-average semantic discriminability receive a lower prior probability of requiring each attribute, while items with lower-than-average discriminability receive a higher prior probability. This formalises the assumption that semantically clear items tend to target fewer skills, with the standardisation ensuring that the influence of $\lambda$ is calibrated symmetrically around the baseline sparsity level $\theta$ and is not sensitive to the scale of the raw $\tau$ values. When $\lambda = 0$, the prior ignores the text information entirely and reduces to the Bernoulli $Q$-prior formulation of \cite{ma2026dynamic}. One may allow the text influence to vary across items by specifying a parameter $\lambda_j$ for each item $j$. In the present study we focus on a parsimonious specification and set $\lambda_j = \lambda$ for all $j$, treating the text influence as constant across items. Estimation of item-specific $\lambda_j$ is a natural extension.

We place a $\mathcal{N}(0,\sigma_\lambda^2)$ prior on $\lambda$, with $\sigma_\lambda$ chosen to allow a moderate influence of the text signal on the logit scale. The value of $\sigma_\lambda$ is specified and examined in the empirical study. Given that $\tau_j$ is standardised, this calibration ensures that the full range of the text signal can shift prior inclusion probabilities noticeably while leaving the response likelihood as the dominant source of information about the $Q$-matrix. The data can therefore override an uninformative or misleading text signal
when the responses are sufficiently informative.

Identification of the DINA model imposes constraints on the $Q$-matrix that must be respected by the prior. Specifically, the necessary and sufficient conditions for identifiability require that $Q$ contains at least two identity submatrices $I_K$, that each column has at least three entries equal to 1, and that, after excluding $I_K$, the remaining submatrix consists of $K$ mutually distinct column vectors \citep{gu2021sufficient}. These conditions are enforced as hard constraints during sampling: proposed $Q$-matrices that violate them are rejected regardless of their prior probability. The text-informed prior therefore operates within the space of identifiable $Q$-matrices.

This construction offers three properties that make it a principled choice for the present setting. First, the NLP signal enters through the prior on $Q$-matrix row complexity rather than through the measurement model itself, so the interpretability of the CDM framework is fully preserved. Second, because the response likelihood remains central, a misleading text signal can in principle be overridden by the data. Third, the approach is especially well suited to settings where some Q-matrix rows are weakly identified from responses alone: even modest prior information about plausible row complexity can improve the stability of Q-matrix recovery without committing to a fully specified item--attribute mapping.

\subsection{Posterior Inference}

Prior distributions for the item parameters follow \cite{ma2026dynamic}. Guessing and slipping parameters are assigned flat priors,
$g_j, s_j \sim \mathrm{Beta}(1,1)$, initialised from
$\mathrm{Uniform}(0, 0.3)$ to reflect the empirical observation that these parameters rarely exceed $0.3$ in applied settings
\citep{zhang2018modeling, culpepper2016revisiting}. Regression coefficients $\boldsymbol{\beta}$ and $\boldsymbol{\gamma}$ are assigned $\mathcal{N}(0,1)$ priors, with all continuous covariates standardised prior to analysis. The global sparsity parameter $\theta$ in the text-informed prior is assigned a $\mathrm{Beta}(\alpha, \beta)$ hyperprior, allowing the data to inform the overall density of the
$Q$-matrix rather than fixing it in advance. The specific values of $\alpha$ and $\beta$ are chosen to reflect prior beliefs about $Q$-matrix density in the application at hand and are reported in the empirical study.  

Combining these priors with the measurement model and transition
structure, the joint posterior distribution is
\begin{equation}
  P(\mathbf{Q}, \mathbf{g}, \mathbf{s}, \boldsymbol{\beta},
    \boldsymbol{\gamma}, \boldsymbol{\alpha}_{1:T}
    \mid \mathbf{Y}_{1:T}, \mathbf{Z})
  \propto
  P(\mathbf{Y}_{1:T} \mid \mathbf{Q}, \mathbf{g}, \mathbf{s},
    \boldsymbol{\alpha}_{1:T})\,
  P(\boldsymbol{\alpha}_{1:T} \mid \boldsymbol{\beta},
    \boldsymbol{\gamma}, \mathbf{Z})\,
  P(\mathbf{Q} \mid \boldsymbol{\tau})\,
  P(\mathbf{g}, \mathbf{s})\,
  P(\boldsymbol{\beta}, \boldsymbol{\gamma}),
\end{equation}
where $P(\mathbf{Q} \mid \boldsymbol{\tau})$ is the text-informed prior specified in equation~\eqref{eq:lambda_T}, replacing the Bernoulli--Beta prior of \cite{ma2026dynamic}. The posterior is
sampled via MCMC using a custom row-wise Gibbs sampler implemented in
\texttt{nimble} \citep{deValpine2026nimble}, with
proposed $Q$-matrices sampled from the identifiable space. Sentence embeddings for $\boldsymbol{\tau}$
are computed before fitting the model using the \texttt{sentence-transformers} library \citep{reimers2019sentencebert}
and treated as fixed inputs throughout estimation.

\subsection{Extensions}
\label{sec:methods_generalised}

The text signal $\tau_j$ used in the main model is constructed from the semantic contrast between the correct option and the distractors, and operates at the item level. When additional textual information is available, the framework can be extended in two natural directions.

The first is available when each attribute has a textual description. In that case, a text signal can be constructed at the item--attribute level by computing
\begin{equation}
  U_{jk} = \mathrm{sim}(\mathbf{e}^{(q)}_j, \mathbf{e}^{(a)}_k),
\end{equation}
where $\mathbf{e}^{(a)}_k$ is the embedding of the description of attribute $k$. The quantity $U_{jk}$ measures how semantically similar item $j$ is to attribute $k$ specifically, and can serve directly as a predictor for $q_{jk}$ rather than for the entire row. For instance, if attributes are described as vocabulary knowledge, syntactic knowledge, and inferential reasoning, these descriptions can be embedded in the same semantic space as the item texts, allowing $U_{jk}$ to be computed for each item--attribute pair.

The two signals can also be combined. Defining
\begin{equation}
  \tau_{jk}^{*} = a\, U_{jk} + b\, \tau_j,
\end{equation}
with weighting coefficients $a$ and $b$, the prior for $q_{jk}$ is then informed both by how semantically related the item is to the description of attribute $k$ and by the overall semantic discriminability of the item. In the present study, attribute descriptions are not available, so we rely on the item-level signal $\tau_j$. The extensions described here indicate how the framework could be applied more directly to settings where richer textual metadata accompanies the assessment.

\section{Empirical Study}

\subsection{Data}

The empirical analysis used log files from Boost Reading, focusing on two games, \textit{Idiomatica} from the vocabulary skills family and \textit{Debate-a-ball} from the comprehension skill family. Students' responses were observed at two time points (Grades 2 and 3), from 2023 to 2025. Figure \ref{fig:0} illustrates the hierarchical structure of the Boost Reading log files. The left panel presents the full structure, including the 11 skill families, their associated 48 games, as well as levels, questions, and attempts. The right panel highlights the subset used in this study, which focuses on the vocabulary and comprehension skill families and includes one game from each.

In Boost Reading, students progress through levels by answering a sequence of questions presented in a fixed order. In this study, we focus on the question level and treat each individual question as an item with a binary correctness response (correct vs. incorrect). The log files provide both question-level information (e.g., binary correctness and response time) and level-level information (e.g., mastery status, percentage correct, and time elapsed to mastery). For the same skill family, there are multiple games designed to support the reading-related skills, as shown in Figure \ref{fig:0}. In this study, we are interested in \textit{Idiomatica} and \textit{Debate-a-ball}.

\textit{Idiomatica}, from the vocabulary skill family, consists of 18 levels with 6-10 questions per level, in total 138 questions. In this game, the question stem and help text (the context that is presented right after the question stem) were combined to form the text information associated with each item, while the correct answer and distractors were presented as the response options. Students identify, define, and apply idioms by answering riddles to navigate through an enchanted maze and recover the lost language of Figura, a land robbed of its colourful expressions. \textit{Debate-a-ball}, from the comprehension skill family, consists of 8 levels with 9 questions per level. In this game, each item requires students to select an answer and identify the evidence sentence that best supports their choice. In the present study, each item is constructed using the question stem associated with the correct evidence, the correct option, and the distractor options.

To construct the sample, we balanced the trade-off between including more students and including more items from the raw log files. We first identified the high frequency questions with the highest levels of student engagement within each game and grade. According to Figures \ref{fig:debate_topgamesG12} and \ref{fig:idio_topgames} in Appendix, we selected the top 10 items from each game at each time point. We next restricted the sample to students who appeared at both time points and completed all selected items across both games and both time points. This procedure gives a consistent longitudinal cohort of 1,616 students from 2023 to 2025. At each time point, 20 questions (i.e., items) were analyzed: 10 items from \textit{Debate-a-ball} and 10 items from \textit{Idiomatica}. Thus, each student contributed 20 binary item responses at Grade 2 and 20 binary item responses at Grade 3. Although the specific questions were not identical across time points, they were designed to measure the same underlying skills, allowing meaningful comparison of latent skill mastery over time. Details of the data cleaning and preprocessing procedures are provided in the Appendix. The selected questions for each game are also listed in the Appendix.

\begin{figure}
\centering
\caption{The hierarchical structure of the log files. The left column shows the full structure of Boost Reading (skill families, games, levels, questions, and attempts). The right column highlights the subset selected for analysis, including two games, their level and question structure, and the corresponding log-based variables used in the study.}
\label{fig:0}

\begin{tikzpicture}[
    node distance = 7mm and 18mm,
    cat/.style = {
        rectangle,
        draw,
        rounded corners,
        minimum width = 13em,
        minimum height = 5em,
        text width = 11em,
        align = center,
        font = \small
    },
    var/.style = {
        rectangle,
        draw,
        rounded corners,
        minimum width = 12em,
        minimum height = 2.6em,
        text width = 19em,
        align = center,
        font = \footnotesize
    },
    line/.style = {-{Latex[length=2mm]}}
]

\node[font=\bfseries] (headerL) {Full structure};
\node[font=\bfseries, right=5cm of headerL] (headerR) {Subset for analysis};

\draw[line] (headerL.south) -- ++(0,-5mm);
\draw[line] (headerR.south) -- ++(0,-5mm);

\node[cat, below=of headerL] (skills)
{11 Skill families\\{\scriptsize (Amplify-defined reading-related skills)}};

\node[cat, below=of skills] (games)
{48 Games};

\node[cat, below=of games] (levels)
{Levels\\{\scriptsize (Varies by game)}};

\node[cat, below=of levels] (questions)
{Questions\\{\scriptsize (Fixed order within level)}};

\node[cat, below=of questions] (attempts)
{Attempts\\{\scriptsize (Varies by student)}};

\draw[line] (skills.south) -- (games.north);
\draw[line] (games.south) -- (levels.north);
\draw[line] (levels.south) -- (questions.north);
\draw[line] (questions.south) -- (attempts.north);

\node[var, right=of skills] (varS)
{2 skill families: vocabulary and comprehension};

\node[var, right=of games] (varG)
{2 games: \textit{Debate-a-ball} and \textit{Idiomatica}};

\node[var, right=of levels, yshift=6mm] (varL1)
{\textit{Debate-a-ball}: 8 levels};

\node[var, right=of levels, yshift=-6mm] (varL2)
{\textit{Idiomatica}: 18 levels};

\node[var, right=of questions, yshift=6mm] (varQ1)
{\textit{Debate-a-ball}: 18 questions per level\\144 questions total};

\node[var, right=of questions, yshift=-6mm] (varQ2)
{\textit{Idiomatica}: 6-10 questions per level\\138 questions total};

\node[var, right=of attempts, yshift=6mm] (varA)
{multiple attempts};

\node[var, right=of attempts, yshift=-6mm] (varB)
{Log-based variables\\{\scriptsize (Correctness, scores, elapsed time)}};

\draw[line] (skills.east) -- (varS.west);
\draw[line] (games.east) -- (varG.west);
\draw[line] (levels.east) -- (varL1.west);
\draw[line] (levels.east) -- (varL2.west);
\draw[line] (questions.east) -- (varQ1.west);
\draw[line] (questions.east) -- (varQ2.west);
\draw[line] (attempts.east) -- (varA.west);
\draw[line] (attempts.east) -- (varB.west);

\end{tikzpicture}
\end{figure}

A set of individual covariates was incorporated into the framework. Students' initial reading performance was assessed using the Dynamic Indicators of Basic Early Literacy Skills (DIBELS; \cite{universityoforegon2018dibels}), administered prior to participation in the Boost program. These scores were categorized into benchmark levels and used for initial placement on the Boost Reading platform. The other categorical covariates were race, special educational needs (SEN), English learner status (ELL), and gender. In addition, several continuous behavioral measures derived from Boost Reading platform usage were included, such as average response time, number of attempts, and the number of questions corrected. The descriptive statistics are summarised in Tables \ref{tab:summary_categorical} and \ref{tab:summary_continuous}.

\begin{table}[ht]
\centering
\caption{Summary of Categorical Variables. \\
\textit{Note: Gender includes female (0) and male (1); SEN = 1 if the student has special educational needs, 0 otherwise; ELL = 1 if the student is an English language learner, 0 otherwise. Race was recoded into three categories: White, Asian, and underrepresented minority (URM). The URM category includes students identified as Black or African American, Other, American Indian or Alaska Native, Native Hawaiian or Other Pacific Islander, Multiracial, or not specified). Not specified race accounts for 640 students (39.6\%) and is included in the URM category.}}
\begin{tabular}{ll}
\hline
\textbf{Variable} & \textbf{Summary} \\
\hline
\multicolumn{2}{l}{\textit{Demographic Variables}} \\
\hline
ELL & Not ELL: 1,449 (90.5\%); ELL: 153 (9.5\%)\\
SEN & Non-SEN: 1,499 (93.6\%); SEN: 103 (6.4\%)\\
Gender & Female: 777 (48.1\%); Male: 825 (51.1\%) \\
Race & White: 532 (32.9\%); Asian: 110 (6.8\%) \\
     & URM: 979 (60.3\%) \\
\hline
\multicolumn{2}{l}{\textit{Initial Literacy Ability (from DIBELS scores)}} \\
\hline
Initial Literacy Ability & Above Benchmark: 880 (54.5\%); At Benchmark: 643 (39.8\%); \\
                         & Below Benchmark: 79 (4.9\%); Well Below Benchmark: 14 (0.9\%) \\
\hline
\label{tab:summary_categorical}
\end{tabular}
\end{table}

As shown in Table \ref{tab:summary_categorical}, students were distributed across different benchmark levels on DIBELS, with a notable proportion classified as Above Benchmark or At Benchmark defined by Amplify criteria. This distribution is partly explained by the grade levels targeted by the games included in the study (\textit{Debate-a-ball}: designed for Grades 2–3; \textit{Idiomatica}: designed for Grades 3–5), which may be more accessible to higher performing students. In contrast, students below benchmark begin with below-grade content and may only encounter these games after many hours play. Most students in this study were not English language learners (not ELL) and did not have special education needs (non-SEN). The gender distribution was relatively balanced, with a slightly higher proportion of male students. The sample included students from diverse racial and ethnic backgrounds, including White,  Black or African American, Other, American Indian or Alaska Native, Native Hawaiian or Other Pacific Islander, Multiracial and Asian. Table \ref{tab:summary_continuous} presents the summary statistics for the log-based variables. These variables were derived from the log files of the 20 selected items administered in Grade 2, based on the sample used in this study.

\begin{table}[ht]
\centering
\caption{Summary of Continuous Variables.\\
\textit{Note: nra = number of reattempts; nlm = number of correctly answered questions; rt = response time. }}
\label{tab:summary_continuous}
\begin{tabular}{lccccc}
\hline
\textbf{Covariate} & \textbf{Mean} & \textbf{SD} & \textbf{Median} & \textbf{(Min, Max)} \\
\hline
Average Attempts (\textit{Debate-a-ball}) & 5.67 & 3.48 & 5 & (3, 41) \\
Average Attempts (\textit{Idiomatica})   & 10.04 & 6.68 & 8 & (5, 88) \\
Correct Answers (\textit{Debate-a-ball})  & 9 & 1.08 & 9 & (4, 10) \\
Correct Answers (\textit{Idiomatica})    & 10 & 0.02 & 10 & (7, 10) \\
Response Time (\textit{Debate-a-ball})    & 5.51 & 1.76 & 5.16 & (3.26, 49.91) \\
Response Time (\textit{Idiomatica})  & 3.29 & 1.50 & 2.97 & (1.06, 28.37) \\
\hline
\end{tabular}
\end{table}

To illustrate the procedure of interacting with the game, we provide examples about three question designs with one correct answer and two distractors from the game \textit{Idiomatica} in Table \ref{tab:example_item1}. To construct the SBERT-based text representation, we first combined the Question and Help Text into a single sequence and encoded it using SBERT. We then computed the similarity between this representation and each response option, including the correct answer and distractors.

\begin{table}[htbp]
\caption{Example three questions from \textit{Idiomatica}.}
\centering
\small
\begin{tabular}{>{\raggedright\arraybackslash}p{0.7cm} 
                >{\raggedright\arraybackslash}p{3.5cm} 
                >{\raggedright\arraybackslash}p{2.5cm} 
                >{\raggedright\arraybackslash}p{2.0cm} 
                >{\raggedright\arraybackslash}p{2.0cm} 
                >{\raggedright\arraybackslash}p{2.0cm}}
\hline
\textbf{Level} & \textbf{Question} & \textbf{Help Text} & \textbf{Correct Answer} & \textbf{Distractor 1} & \textbf{Distractor 2} \\
\hline
1 &
Hey there, do you know any \textit{bookworms}? &
What does \textit{bookworm} mean? &
Yes I do! They are always reading books. &
Um, no...I don't know any worms! &
Yes, books are delicious! \\
\hline
2 &
Every good \textit{bookworm} could use a book. I've got one for you if you're up for it. &
What does \textit{bookworm} mean? &
Sure, I'll take it. I ... love reading books. &
Sure, I'll take it. I ... could use a snack. &
Sure, I'll take it. I ... could use a nap. \\
\hline
3 &
I hope you'll enjoy reading this book—after you get out of this dark maze, that is. &
Which one means you like books? &
I'm sure I will. I'm a real... bookworm. &
I'm sure I will. I'm a real ... bookend. &
I'm sure I will. I'm a real ... glow worm. \\
\hline
\end{tabular}
\label{tab:example_item1}
\end{table}

To illustrate the construction of the text-derived signal, we provide a simple example for item 1 presented in Table \ref{tab:example_item1} using the first three dimensions of the embeddings. Let \begin{align*} 
\mathbf{e}^{(q)}_1 &= (-0.00559068,\; 0.00106812,\; 0.02144735),\\
\mathbf{e}^{(c)}_1 &= (0.07255946,\; -0.02993354,\; 0.04118648),\\
\mathbf{e}^{(d_1)}_1 &= (-0.0366895,\; -0.01831677,\; 0.05275081),\\
\mathbf{e}^{(d_2)}_1 &= (-0.0213028,\; -0.04499649,\; 0.00077056).
\end{align*}
Using cosine similarity,
$
\mathrm{sim}(\mathbf{u},\mathbf{v})=
\frac{\mathbf{u}^{\top}\mathbf{v}}
{\|\mathbf{u}\|\,\|\mathbf{v}\|},
$
we computed
$$
S_1^{+} = \mathrm{sim}\!\left(\mathbf{e}^{(q)}_1,\mathbf{e}^{(c)}_1\right),
S_1^{-} = \frac{1}{2}\left[
\mathrm{sim}\!\left(\mathbf{e}^{(q)}_1,\mathbf{e}^{(d_1)}_1\right)
+\mathrm{sim}\!\left(\mathbf{e}^{(q)}_1,\mathbf{e}^{(d_2)}_1\right)
\right].
$$
The final text-derived quantity was then calculated using
$\tau_1 = S_1^{+} - S_1^{-} = -0.134.$ For illustration, only the first three dimensions are shown here, while the actual computation uses the full embedding vectors. The distribution of the item pool is shown in Figure \ref{fig:Tdistri}, standardizing $\tau$ to have mean 0 and variance 1 in Figure \ref{fig:Tdistri_standard}. Because $\tau_j$ is constructed based on item-specific structures, its scale may vary across items due to differences in construction. To ensure comparability across items constructed in different ways, we standardize $\tau$ to have mean 0 and variance 1, to reduce sensitivity of the proposed framework to construction-induced scale differences.

\subsection{Model Specification for the Empirical Study}

The $Q$-matrices at the two time points, denoted by $Q_1$ and $Q_2$, were treated as unknown and estimated jointly with the latent attribute profiles, item parameters, initial mastery effects, and transition effects. To incorporate text information, we used an item-level prior specification of the form
$$
q_{jk}^{(t)} \sim \mathrm{Bernoulli}(\pi_j^{(t)}), \qquad t=1,2,
$$
with
$$
\mathrm{logit}(\pi_j^{(t)}) = \mathrm{logit}(\theta) - \lambda \tau_j^{(t)},
$$
where $\theta$ is a global sparsity parameter and $\lambda$ controls the global strength of the text-informed prior. Under this specification, the same item-level text-derived quantity influences the prior probabilities of the item-attribute indicators for that item at a given time point.

We specified a $\mathrm{Beta}(24,6)$ prior for $\theta$ in equation \eqref{eq:lambda_T}, which has a mean of $0.8$ and a variance of approximately $0.0052$, placing substantial prior mass between 0.65 and 0.92. Although the model is conceptually motivated by $\lambda \geq 0$, we place a $\mathcal{N}(0, 0.5^2)$ prior on $\lambda$ rather than enforcing a hard positivity constraint, allowing the data to inform the direction of the effect. When $\lambda = 0$, the model reduces to the baseline specification without text information. For the guessing and slipping parameters, we used non-informative priors (i.e., flat priors) following \cite{ma2026dynamic}:
$$g_{j,t} \sim \mathrm{Beta}(1, 1),
s_{j,t} \sim \mathrm{Beta}(1, 1). $$
For the regression coefficients in the attribute and transition models (i.e., $\boldsymbol\beta_Z$, $\gamma_{01}$, and $\gamma_{10}$), we assigned independent normal priors:
$\boldsymbol\beta_Z, \gamma_{01}, \gamma_{10} \sim \mathcal{N}(0, 1).$

\subsection{Estimation}

The empirical model was estimated in \texttt{nimble} using Markov chain Monte Carlo (MCMC). Because the Q-matrix was treated as unknown, we implemented a custom row-wise Gibbs sampler to update each item-specific Q-vector jointly. For $K=2$, the candidate non-zero row patterns of Q-matrix were $(0,1)$, $(1,0)$, and $(1,1)$. At each MCMC update, candidate row patterns were evaluated subject to identifiability constraints. In particular, zero rows were excluded, each attribute was required to appear in a sufficient number of items, and at least one pure item was required for each attribute. Only candidate rows satisfying these constraints were assigned positive posterior weight. The remaining model parameters, including item parameters, regression coefficients, transition parameters, and latent attribute states, were sampled within the same MCMC framework. Multiple chains were run from different initial values, and convergence was assessed using standard diagnostics based on the potential scale reduction factor.

\subsection{Empirical results}

We applied the proposed model to the dataset and assessed convergence following \cite{vehtari2021rank}. From 30,000 total iterations (3 chains with 10,000 for each), the first half were discarded as warm-up. Diagnostics indicated the maximum potential scale reduction factor ($\hat{R}$) with 1.01. The average effective sample size (ESS) was 1160 with minimum 696. Running time for the empirical analysis was approximately 43 minutes, conducted on a MacBook Pro (13-inch, M1, 2020) equipped with an Apple M1 chip (8-core: 4 performance and 4 efficiency cores) and 16 GB of unified memory.

\begin{table}[ht]
\centering
\caption{Comparison of estimated $Q$-matrices, guessing ($g$), and slipping ($s$) parameters under the Baseline model and the text-prior model. Entries in italics indicate differences between the two models.}
\label{tab:qmatrix_compare_text}
\scriptsize
\resizebox{\textwidth}{!}{
\begin{tabular}{c|cccc|cccc|cccc|cccc}
\toprule
\multirow{3}{*}{Item}
& \multicolumn{8}{c|}{\textbf{Time 1}}
& \multicolumn{8}{c}{\textbf{Time 2}} \\
\cmidrule(lr){2-9} \cmidrule(lr){10-17}
& \multicolumn{4}{c|}{\textbf{Baseline}}
& \multicolumn{4}{c|}{\textbf{With text}}
& \multicolumn{4}{c|}{\textbf{Baseline}}
& \multicolumn{4}{c}{\textbf{With text}} \\
\cmidrule(lr){2-5} \cmidrule(lr){6-9}
\cmidrule(lr){10-13} \cmidrule(lr){14-17}
& $A_1$ & $A_2$ & $g$ & $s$
& $A_1$ & $A_2$ & $g$ & $s$
& $A_1$ & $A_2$ & $g$ & $s$
& $A_1$ & $A_2$ & $g$ & $s$ \\
\midrule
1  & 1 & 0 & 0.208 & 0.020 & 1 & 0 & 0.209 & 0.019 & 1 & 0 & 0.442 & 0.157 & 1 & 0 & 0.432 & 0.158 \\
2  & 1 & 0 & 0.265 & 0.024 & 1 & 0 & 0.264 & 0.024 & 1 & 0 & 0.476 & 0.154 & 1 & 0 & 0.473 & 0.151 \\
3  & 1 & 0 & 0.243 & 0.003 & 1 & 0 & 0.244 & 0.013 & 1 & 0 & 0.456 & 0.143 & 1 & 0 & 0.454 & 0.142 \\
4  & 1 & 0 & 0.278 & 0.008 & 1 & 0 & 0.278 & 0.009 & 1 & 0 & 0.491 & 0.142 & 1 & 0 & 0.491 & 0.143 \\
5  & 1 & 0 & 0.315 & 0.005 & 1 & 0 & 0.316 & 0.005 & 1 & 0 & 0.512 & 0.133 & 1 & 0 & 0.502 & 0.134 \\
6  & 1 & 0 & 0.339 & 0.007 & 1 & 0 & 0.338 & 0.007 & 1 & 0 & 0.450 & 0.126 & 1 & 0 & 0.450 & 0.126 \\
7  & 1 & 0 & 0.478 & 0.275 & 1 & 0 & 0.479 & 0.274 & 0 & 1 & 0.747 & 0.205 & 0 & 1 & 0.748 & 0.206 \\
8  & 1 & 0 & 0.496 & 0.249 & 1 & 0 & 0.497 & 0.248 & 1 & 0 & 0.728 & 0.180 & 1 & 0 & 0.724 & 0.187 \\
9  & 1 & 0 & 0.546 & 0.241 & 1 & 0 & 0.545 & 0.240 & 1 & 0 & 0.763 & 0.210 & 1 & 0 & 0.762 & 0.209 \\
10 & 1 & 0 & 0.553 & 0.251 & 1 & \textit{0} & 0.553 & 0.252 & 1 & \textit{0} & 0.739 & 0.202 & 1 & \textit{1} & 0.635 & 0.301 \\
11 & 0 & 1 & 0.250 & 0.148 & 0 & 1 & 0.252 & 0.141 & 1 & 1 & 0.426 & 0.073 & 1 & 1 & 0.436 & 0.074 \\
12 & 0 & 1 & 0.244 & 0.126 & 0 & 1 & 0.246 & 0.121 & 1 & 1 & 0.485 & 0.112 & 1 & 1 & 0.484 & 0.112 \\
13 & 0 & 1 & 0.368 & 0.090 & 0 & 1 & 0.368 & 0.088 & \textit{1} & 1 & 0.557 & 0.027 & \textit{0} & 1 & 0.568 & 0.024 \\
14 & 0 & 1 & 0.384 & 0.053 & 0 & 1 & 0.385 & 0.052 & 1 & 1 & 0.398 & 0.039 & 1 & 1 & 0.399 & 0.040 \\
15 & 0 & 1 & 0.212 & 0.173 & 0 & 1 & 0.213 & 0.172 & 0 & 1 & 0.393 & 0.045 & 0 & 1 & 0.387 & 0.046\\
16 & 0 & 1 & 0.338 & 0.393 & 0 & 1 & 0.339 & 0.392 & \textit{0} & 1 & 0.314 & 0.244 & \textit{1} & 1 & 0.316 & 0.242 \\
17 & 0 & 1 & 0.446 & 0.348 & 0 & 1 & 0.446 & 0.358 & 1 & 1 & 0.316 & 0.099 & 1 & 1 & 0.315 & 0.099 \\
18 & 0 & 1 & 0.489 & 0.208 & 0 & 1 & 0.491 & 0.206 & 0 & 1 & 0.341 & 0.063 & 0 & 1 & 0.342 & 0.063 \\
19 & 0 & 1 & 0.595 & 0.171 & 0 & 1 & 0.585 & 0.170 & 1 & 1 & 0.353 & 0.018 & 1 & 1 & 0.343 & 0.028 \\
20 & 0 & 1 & 0.641 & 0.125 & 0 & 1 & 0.642 & 0.126 & \textit{0} & 1 & 0.352 & 0.419 & \textit{1} & 1 & 0.338 & 0.432 \\
\bottomrule
\end{tabular}}
\end{table}

Table \ref{tab:qmatrix_compare_text} compares the estimated $Q$-matrices and item parameters under the baseline and text-informed models. The estimated Q-matrix structure was largely consistent across the two models. At Time 1, attribute assignments were identical across all items. At Time 2, the two models differed only for Items 10, 13, 16, and 20, which correspond to items with comparatively less stable posterior $Q$-row distributions. The posterior mean of $\lambda$ was 0.128 with a standard deviation of 0.429. The positive posterior mean is thus consistent with our assumption that higher semantic discriminability is associated with sparser
$Q$-matrix rows, providing empirical support for the construction of $\tau_j$. At the same time, the modest magnitude indicates that the response data in this application were sufficiently informative to identify the $Q$-matrix structure without heavy reliance on the text prior, which is not unexpected: with $K=2$ and $J=20$, the identifiable space of
$Q$-matrices is rather constrained, leaving limited room for the prior to shift posterior mass. The text prior nevertheless provided consistent, if modest, regularisation for the less stable items. Its more pronounced benefit in lower-information settings is illustrated in the simulation study. 

Using the text-based prior model, Table \ref{tab:profile_distribution_attribute_transition_matrix} presents the estimated transition matrix of attribute profiles. At Time 1, more students mastered \textit{Idiomatica} than \textit{Debate-a-ball} alone. From Time 1 to Time 2, many students with \textit{Idiomatica} only mastery transitioned to full mastery of both games. In contrast, relatively few students achieved \textit{Debate-a-ball} mastery without also mastering \textit{Idiomatica}. 

Table \ref{tab:betaz_gamma01_significant} presents the posterior mean odds ratios for initial mastery ($\beta_z$) and transition probabilities ($\gamma_{01}$) by attribute $K$. Covariates include log-based variables, demographics, and initial literacy ability (see Tables \ref{tab:summary_categorical} and \ref{tab:summary_continuous}), with only statistically significant effects reported in \ref{tab:betaz_gamma01_significant}. Detailed posterior means and 95\% confidence intervals of the odds ratios are reported in Tables \ref{tab:betaZ_OR_part1}–\ref{tab:gamma01_OR_part2}.

For initial mastery ($\beta_z$), a greater number of reattempts in \textit{Idiomatica} was positively associated with mastery of the first attribute ($K = 1$, i.e., the vocabulary skill), and higher initial literacy ability was negatively associated. For the second attribute ($K = 2$, i.e., the comprehension skill), higher initial literacy ability was positively associated with mastery, while one of the Race categories (Asian group) was negatively associated with this. For transitions ($\gamma_{01}$), longer response time in \textit{Debate-a-ball} and higher initial literacy ability were negatively associated with transitioning to mastery for the vocabulary skill. In contrast, for the comprehension skill, higher initial literacy ability and one of the Race categories (Asian group) were positively associated with transition.

\begin{table}[ht]
\centering
\caption{Transition matrix of attribute profiles from Time 1 (rows) to Time 2 (columns) for 1,616 students. Each element shows the number of students (proportion). Row sums represent Time 1 distributions; column sums represent Time 2 distributions. Profile labels: 00 = no mastery, 10 = vocabulary skill only, 01 = comprehension skill only, 11 = mastery of both skills.}
\label{tab:profile_distribution_attribute_transition_matrix}
\setlength{\tabcolsep}{2.5pt}
\begin{tabular}{llcccc|c}
\toprule
 & & \multicolumn{4}{c|}{\textbf{Time 2}} & \textbf{Totals} \\
\cmidrule(lr){3-6}
 & & 00 & 10 & 01 & 11 & \\
\midrule
\multirow{4}{*}{\textbf{Time 1}} 
 & 00 & 66(4.08\%) & 64(3.96\%) & 5(0.31\%) & 76(4.70\%) & 211(13.05\%) \\
 & 10 & 117(7.24\%) & 150(9.28\%) & 12(0.74\%) & 227(14.05\%) & 506(31.31\%) \\
 & 01 & 37(2.29\%) & 42(2.60\%) & 4(0.25\%) & 102(6.31\%) & 185(11.45\%) \\
 & 11 & 137(8.48\%) & 212(13.12\%) & 12(0.74\%) & 353(21.84\%) & 714(44.19\%) \\
\midrule
\multicolumn{2}{l}{\textbf{Totals}} 
& 357(22.09\%) & 468(28.96\%) & 33(2.04\%) & 758(46.91\%) & 1616(100\%) \\
\bottomrule
\end{tabular}
\end{table}

\begin{table}[htbp]
\centering
\caption{Significant covariates for $\beta_z$ (initial mastery) and $\gamma_{01}$ by attribute ($K$). Only covariates with statistically significant odds ratios (OR) are shown.\\
\textit{Note}: $K$ = attribute; NRA = average number of attempts; RT = average response time; ILA = initial literacy ability (at benchmark: reference level); WB = well below benchmark; BB = below benchmark; AB = above benchmark; Asian is coded relative to the White reference group (one of the Race categories).}
\label{tab:betaz_gamma01_significant}
\begin{tabular}{ccc|ccc}
\toprule
\multicolumn{3}{c|}{Initial mastery} & \multicolumn{3}{c}{Transition probability} \\
\multicolumn{3}{c|}{$\beta_z$} & \multicolumn{3}{c}{$\gamma_{01}$} \\
$K$ & Covariates & OR & $K$ & Covariates & OR \\
\midrule
1 & NRA \textit{Idiomatica} & 2.085 & 1 & RT \textit{Debate-a-ball} & 0.305 \\
1 & ILA-WB & 0.015 & 1 & ILA-BB & 0.817 \\
1 & ILA-BB & 0.002 & 2 & ILA-WB & 1.619 \\
1 & ILA-AB & 1.271 & 2 & Asian & 6.942 \\
2 & ILA-WB & 2.407 &   &   &   \\
2 & Asian & 0.251 &   &   &   \\
\bottomrule
\end{tabular}
\end{table}

\begin{sidewaystable}[htbp]
\centering
\caption{
Posterior means of odds ratios (OR) for covariate effects $\beta_z$ by attribute ($K$), with 95\% credible intervals (CI). Statistically significant results (CI excluding 1) are shown in \textbf{bold}. Part 1 of 2.\\
\textit{Note}: rt = average response time; nlm = number of questions correct; n\_attempts = average number of attempts; gender = female (0), male (1). idio = \textit{Idiomatica}; debate = \textit{Debate-a-ball}.
}
\label{tab:betaZ_OR_part1}
\begin{tabularx}{\linewidth}{ll*{7}{>{\centering\arraybackslash}X}}
\toprule
\textbf{K} & \textbf{Measure} & \textbf{rt debate} & \textbf{rt idio} & \textbf{nlm debate} & \textbf{nlm idio} & \textbf{n\_attempts debate} & \textbf{n\_attempts idio} & \textbf{gender} \\
\midrule
1 & OR & 0.880 & 1.069 & 1.060 & 0.969 & 0.956 & \textbf{2.085} & 2.029\\
  & CI & (0.723, 1.044) & (0.853, 1.339) & (0.913, 1.241) & (0.806, 1.161) & (0.811, 1.119) & (1.734, 2.525) & (0.872, 8.841) \\
\addlinespace
2 & OR & 1.006 & 0.991 & 1.002 & 0.971 & 0.524 & 1.402 & 1.008 \\
  & CI & (0.137, 7.149) & (0.137, 6.872) & (0.141, 6.546) & (0.139, 7.059) & (0.131, 2.218) & (0.404, 5.416) & (0.484, 2.161) \\
\bottomrule
\end{tabularx}
\end{sidewaystable}

\begin{sidewaystable}[htbp]
\centering
\caption{
Posterior means of odds ratios (OR) for covariate effects $\beta_z$ by attribute ($K$), with 95\% credible intervals (CI). Statistically significant results (CI excluding 1) are shown in \textbf{bold}. Part 2 of 2.\\
\textit{Note}: SEN = 1 if the student has special educational needs, 0 otherwise; ELL = 1 if the student is an English language learner, 0 otherwise. ILA = initial literacy ability; WB = well below benchmark, BB = below benchmark, and AB = above benchmark. Group denotes engagement group, with group 5 as the reference category. Race was recoded into three categories: White, Asian, and underrepresented minority (URM). The URM category includes students identified as Black or African American, Other, American Indian or Alaska Native, Native Hawaiian or Other Pacific Islander, Multiracial, or not specified.
}
\label{tab:betaZ_OR_part2}
\begin{tabularx}{\linewidth}{ll*{7}{>{\centering\arraybackslash}X}}
\toprule
\textbf{K} & \textbf{Measure} & \textbf{SEN (1=yes)} & \textbf{ELL (1=yes)} & \textbf{Benchmark-WB} & \textbf{Benchmark-BB} & \textbf{Benchmark-AB} & \textbf{Race (Asian)} & \textbf{Race (URM)} \\
\midrule
1 & OR & 2.135 & 0.930 & \textbf{0.015} & \textbf{0.002} & \textbf{1.271} &1.018 & 0.993\\
  & CI & (0.722, 9.194) & (0.753, 1.133) & (0.009, 0.026) & (0.001, 0.004) & (1.028, 1.571) & (0.134, 7.304) & (0.132, 7.370) \\
\addlinespace
2 & OR & 0.971 & 0.879 & \textbf{2.407} & 0.998 & 1.017 & \textbf{0.251} & 1.281 \\
  & CI & (0.495, 1.913) & (0.629, 1.253) & (1.685, 3.451) & (0.120, 7.586) & (0.132, 6.677) & (0.093, 0.659) & (0.461, 4.101) \\
\bottomrule
\end{tabularx}
\end{sidewaystable}

\begin{sidewaystable}[htbp]
\centering
\caption{
Posterior means of odds ratios (OR) for $\gamma_{01}$ by attribute ($K$), with 95\% credible intervals (CI). Statistically significant results (CI excluding 1) are shown in \textbf{bold}. Part 1 of 2.\\
\textit{Note}: $K$ = attribute; rt = average response time; nlm = number of questions correct; n\_attempts = average number of attempts; debate = \textit{Debate-a-ball} game; idio = \textit{Idiomatica} game; gender = female (0), male (1).
}
\label{tab:gamma01_OR_part1}
\begin{tabularx}{\linewidth}{ll*{7}{>{\centering\arraybackslash}X}}
\toprule
\textbf{K} & \textbf{Measure} & \textbf{rt debate} & \textbf{rt idio} & \textbf{nlm debate} & \textbf{nlm idio} & \textbf{n\_attempts debate} & \textbf{n\_attempts idio} & \textbf{gender} \\
\midrule
1 & OR  & \textbf{0.305} & 1.406 & 1.047 & 0.946 & 0.934 & 1.136 & 1.120 \\
  & CI  & (0.137, 0.712) & (0.924, 2.403) & (0.857, 1.283) & (0.716, 1.282) & (0.751, 1.153) & (0.861, 1.412) & (0.952, 1.334)\\
\addlinespace
2 & OR  & 0.980 & 0.992 & 1.027 & 1.014 & 1.469 & 0.841 & 1.506 \\
  & CI  & (0.148, 6.741) & (0.144, 6.707) & (0.152, 6.984) & (0.139, 6.813) & (0.356, 7.226) & (0.174, 3.633) & (0.573, 5.616) \\
\bottomrule
\end{tabularx}
\end{sidewaystable}

\begin{sidewaystable}[htbp]
\centering
\caption{
Posterior means of odds ratios (OR) for $\gamma_{01}$ by attribute ($K$), with 95\% credible intervals (CI). Statistically significant results (CI excluding 1) are shown in \textbf{bold}. Part 2 of 2.\\
\textit{Note}: $K$ = attribute; SEN = 1 if the student has special educational needs, 0 otherwise; ELL = 1 if the student is an English language learner, 0 otherwise. ILA = initial literacy ability; WB = well below benchmark, BB = below benchmark, and AB = above benchmark. Group denotes engagement group, with group 5 as the reference category. Race was recoded into three categories: White, Asian, and underrepresented minority (URM).
}
\label{tab:gamma01_OR_part2}
\begin{tabularx}{\linewidth}{ll*{7}{>{\centering\arraybackslash}X}}
\toprule
\textbf{K} & \textbf{Measure} & \textbf{SEN (1=yes)} & \textbf{ELL (1=yes)} & \textbf{Benchmark-WB} & \textbf{Benchmark-BB} & \textbf{Benchmark-AB} & \textbf{Race (Asian)} & \textbf{Race (URM)} \\
\midrule
1 & OR  & 0.976 & 0.469 & 0.862 & \textbf{0.817} & 1.216 & 1.015 & 0.987 \\
  & CI  & (0.144, 6.728) & (0.112, 1.028) & (0.695, 1.056) & (0.687, 0.963) & (0.952, 1.717) & (0.851, 1.209) & (0.150, 6.938) \\
\addlinespace
2 & OR  & 1.781 & 1.542 & \textbf{1.619} & 1.018 & 0.995 & \textbf{6.942} & 1.984 \\
  & CI  & (0.846, 3.637) & (0.867, 2.992) & (1.095, 2.395) & (0.137, 6.890) & (0.142, 7.298) & (1.055, 30.728) & (0.847, 4.810) \\
\bottomrule
\end{tabularx}
\end{sidewaystable}

\section{Simulations}

\subsection{Simulation Design}

The simulation study was designed not only to reflect aspects of the empirical setting but also to evaluate the model under a more general and challenging scenario. We considered a dynamic CDM with two attributes measured across two time points. Sample size varied across conditions $N \in \{800, 1600, 2400\}$, and the number of items administered at each time point varied as $J \in \{10, 20, 30\}$. The true $Q$-matrices used under each simulation condition are provided in Table \ref{tab:qmatrix_j30_t1_t2} in Appendix. The prior for $\theta$ is specified as $\mathrm{Beta}(6,4)$, with a prior mean of $0.6$. The prior for $\lambda$ is specified as $\mathcal{N}(0, 0.5^2)$.

In the simulation study, the item-level text-derived quantity $\tau_j$ was generated from the empirical distribution of observed text-based values. This choice was motivated by empirical evidence that the observed distribution of $\tau_j$ does not follow e.g., a normal distribution (see Appendix Figure \ref{fig:Tdistri}). To preserve the shape of the empirical distribution, we adopted a nonparametric sampling strategy based on kernel density estimation (KDE). Let $\{\tau^{\mathrm{empirical}}_1,\ldots,\tau^{\mathrm{empirical}}_M\}$ denote the empirical text-derived values obtained from the full item pool. A kernel density estimator was constructed as
$$
\hat f_h(t)=\frac{1}{Mh}\sum_{m=1}^M K\!\left(\frac{t-\tau^{\mathrm{empirical}}_m}{h}\right),
$$
where $K(\cdot)$ is a Gaussian kernel and $h$ is the bandwidth selected using the normal reference rule \citep{silverman1986density}. Simulated values of $\tau$ were then generated by sampling from the estimated density $\hat f_h(t)$.

For each simulation replication and each time point, we generated a vector of item-level text quantities,
$$
\tau_t = (\tau_{1t},\ldots,\tau_{Jt}),
$$
by drawing $J$ samples from the estimated density. This approach allows the simulated text signals to preserve the skewness and variability observed in the real data.

We assessed model performance using both parameter estimation and classification accuracy, following \citet{ma2026dynamic}. For item parameters and regression coefficients in the attribute and transition models, mean absolute error (MAE) and root mean square error (RMSE) were computed across replications. Classification performance was evaluated using the profile agreement rate (PAR) and attribute agreement rates (AAR). For the $Q$-matrix, posterior samples were obtained from the MCMC output. For each item $j$, the posterior probabilities of candidate attribute patterns were estimated based on their frequencies in the MCMC samples. A row-wise maximum a posteriori (MAP) rule was then used to obtain a point estimate of the $Q$-matrix by selecting, for each item, the attribute pattern with the highest posterior probability. We used classification accuracy (ACC), false positive rate (FPR) and false negative rate (FNR) to evaluate the performance of $Q$-matrix recovery. Additionally, we computed posterior inclusion probabilities (PIP) for each entry $q_{jk},$ defined as the posterior mean, which represents the probability that $q_{jk} = 1$. To evaluate $Q$-matrix recovery, we summarized the PIP values separately for true and false entries. Specifically, we computed the average PIP over entries where $q_{jk} = 1$ and where $q_{jk} = 0$, reflecting the model’s ability to assign high posterior probability to true associations and low probability to false ones.

\subsection{Simulation Results}

For each simulation condition, we conducted 100 independent replications. The model estimation was implemented using 3 independent Markov chains per replication, each initialized with different starting values to ensure broad coverage of the parameter space. Each chain consisted of 3,000 burn-in iterations, followed by 3,000 monitored iterations for posterior inference. Convergence was assessed using the potential scale reduction factor ($\hat{R}$), with all parameters having $\hat{R}$ values below 1.1. To quantify simulation uncertainty, we applied a nonparametric bootstrap procedure to the 100 replications. Specifically, 1,000 bootstrap samples were drawn with replacement, and standard errors for all performance metrics were derived from these bootstrap distributions. Computational time increased with both sample size and test length, with average runtime per replication ranging from 44 to 100 minutes across conditions.

Under most simulation conditions, both models achieved near-perfect $Q$-matrix recovery. The conditions offering the clearest comparison are those with $J=30$, particularly at smaller sample sizes, where item-level identification is weakest. As shown in Table \ref{tab:q_recovery}, $Q$-matrix recovery improved with increasing sample size ($N$) and number of items ($J$) for both models. The text-prior model generally performed comparably to or slightly better than the baseline in these more challenging settings, often showing lower false positive and false negative rates together with higher mean PIP values for true entries and lower mean PIP values for false entries. Bootstrap standard errors were small across all conditions, indicating stable estimation.

\begin{landscape}
\begin{table}[!h]
\centering
\caption{Recovery of the Q-matrix across time points ($T$) under varying sample sizes ($N$) and numbers of items ($J$), comparing the Baseline model and the Text-prior model. Values are reported as mean (bootstrap SE). The reported metrics include classification accuracy (ACC), false positive rate (FPR), false negative rate (FNR), and mean posterior inclusion probability for true Q-matrix entries (PIP true), and mean posterior inclusion probability for false Q-matrix entries (PIP false).}
\label{tab:q_recovery}
\footnotesize
\setlength{\tabcolsep}{1.5pt}
\begin{tabular}{ccc|ccccc|ccccc}
\toprule
\multicolumn{3}{c|}{} & \multicolumn{5}{c|}{Baseline model} & \multicolumn{5}{c}{Text-prior model} \\
\cmidrule(lr){4-8} \cmidrule(lr){9-13}
$N$ & $J$ & $T$ & ACC & FPR & FNR & PIP (true) & PIP (false) & ACC & FPR & FNR & PIP (true) & PIP (false) \\
\midrule
800  & 10 & 1 & 1.000 (0.000) & 0.000 (0.000) & 0.000 (0.000) & 1.000 (0.000) & 0.000 (0.000) & 1.000 (0.000) & 0.000 (0.000) & 0.000 (0.000) & 1.000 (0.000) & 0.000 (0.000) \\
     &    & 2 & 1.000 (0.000) & 0.000 (0.000) & 0.000 (0.000) & 1.000 (0.000) & 0.000 (0.000) & 1.000 (0.000) & 0.000 (0.000) & 0.000 (0.000) & 1.000 (0.000) & 0.000 (0.000) \\
     & 20 & 1 & 1.000 (0.000) & 0.000 (0.000) & 0.000 (0.000) & 1.000 (0.000) & 0.000 (0.000) & 1.000 (0.000) & 0.000 (0.000) & 0.000 (0.000) & 1.000 (0.000) & 0.000 (0.000) \\
     &    & 2 & 1.000 (0.000) & 0.000 (0.000) & 0.000 (0.000) & 1.000 (0.000) & 0.000 (0.000) & 1.000 (0.000) & 0.000 (0.000) & 0.000 (0.000) & 1.000 (0.000) & 0.000 (0.000) \\
     & 30 & 1 & 0.847 (0.141) & 0.200 (0.182) & 0.124 (0.109) & 0.834 (0.070) & 0.267 (0.113) & 1.000 (0.000) & 0.000 (0.000) & 0.000 (0.000) & 0.959 (0.038) & 0.067 (0.055) \\
     &    & 2 & 1.000 (0.000) & 0.000 (0.000) & 0.000 (0.000) & 1.000 (0.000) & 0.026 (0.023) & 1.000 (0.000) & 0.000 (0.000) & 0.000 (0.000) & 1.000 (0.000) & 0.000 (0.000) \\
\addlinespace[0.3em]
1600 & 10 & 1 & 1.000 (0.000) & 0.000 (0.000) & 0.000 (0.000) & 1.000 (0.000) & 0.000 (0.000) & 1.000 (0.000) & 0.000 (0.000) & 0.000 (0.000) & 1.000 (0.000) & 0.000 (0.000) \\
     &    & 2 & 1.000 (0.000) & 0.000 (0.000) & 0.000 (0.000) & 1.000 (0.000) & 0.000 (0.000) & 1.000 (0.000) & 0.000 (0.000) & 0.000 (0.000) & 1.000 (0.000) & 0.000 (0.000) \\
     & 20 & 1 & 1.000 (0.000) & 0.000 (0.000) & 0.000 (0.000) & 0.940 (0.029) & 0.100 (0.048) & 1.000 (0.000) & 0.000 (0.000) & 0.000 (0.000) & 0.991 (0.009) & 0.031 (0.031) \\
     &    & 2 & 1.000 (0.000) & 0.000 (0.000) & 0.000 (0.000) & 1.000 (0.000) & 0.000 (0.000) & 1.000 (0.000) & 0.000 (0.000) & 0.000 (0.000) & 1.000 (0.000) & 0.000 (0.000) \\
     & 30 & 1 & 0.936 (0.063) & 0.083 (0.079) & 0.052 (0.049) & 0.966 (0.034) & 0.056 (0.052) & 0.936 (0.062) & 0.083 (0.080) & 0.052 (0.050) & 0.931 (0.049) & 0.111 (0.079) \\
     &    & 2 & 0.993 (0.007) & 0.032 (0.030) & 0.000 (0.000) & 1.000 (0.000) & 0.026 (0.025) & 1.000 (0.000) & 0.000 (0.000) & 0.000 (0.000) & 0.992 (0.007) & 0.028 (0.026) \\
\addlinespace[0.3em]
2400 & 10 & 1 & 1.000 (0.000) & 0.000 (0.000) & 0.000 (0.000) & 1.000 (0.000) & 0.000 (0.000) & 1.000 (0.000) & 0.000 (0.000) & 0.000 (0.000) & 1.000 (0.000) & 0.000 (0.000) \\
     &    & 2 & 1.000 (0.000) & 0.000 (0.000) & 0.000 (0.000) & 1.000 (0.000) & 0.000 (0.000) & 1.000 (0.000) & 0.000 (0.000) & 0.000 (0.000) & 1.000 (0.000) & 0.000 (0.000) \\
     & 20 & 1 & 1.000 (0.000) & 0.000 (0.000) & 0.000 (0.000) & 1.000 (0.000) & 0.000 (0.000) & 1.000 (0.000) & 0.000 (0.000) & 0.000 (0.000) & 0.983 (0.016) & 0.028 (0.027) \\
     &    & 2 & 1.000 (0.000) & 0.000 (0.000) & 0.000 (0.000) & 1.000 (0.000) & 0.000 (0.000) & 1.000 (0.000) & 0.000 (0.000) & 0.000 (0.000) & 0.995 (0.005) & 0.028 (0.026) \\
     & 30 & 1 & 0.910 (0.059) & 0.118 (0.074) & 0.073 (0.046) & 0.939 (0.041) & 0.098 (0.067) & 0.955 (0.043) & 0.059 (0.056) & 0.037 (0.036) & 0.927 (0.040) & 0.118 (0.062) \\
     &    & 2 & 0.975 (0.025) & 0.059 (0.056) & 0.016 (0.016) & 0.989 (0.011) & 0.039 (0.039) & 0.955 (0.031) & 0.118 (0.079) & 0.025 (0.017) & 0.978 (0.013) & 0.107 (0.051) \\
\bottomrule
\end{tabular}
\end{table}
\end{landscape}

Table \ref{tab:alpha_recovery} presents PAR and AARs with bootstrap standard errors across conditions. Across all combinations of $N$ and $J$, both PAR and AARs remained high, generally exceeding 0.90 even under more challenging conditions (e.g., smaller $N$ or larger $J$). For a fixed $N$, recovery performance showed slight improvements as $J$ increased, though gains were modest at higher values. As sample size increased, both PAR and AARs improved consistently, often approaching or exceeding 0.98. The text-prior model performed slightly better than the baseline model across most settings.

\begin{table}[!h]
\centering
\caption{Recovery of attribute profiles across time points ($T$) under varying sample sizes ($N$) and numbers of items ($J$), comparing the Baseline model and the Text-prior model. Values are reported as mean (bootstrap SE). The reported metrics include profile agreement rate (PAR) and attribute agreement rates (AAR$_1$, AAR$_2$).}
\label{tab:alpha_recovery}
\footnotesize
\setlength{\tabcolsep}{4pt}
\begin{tabular}{ccc|ccc|ccc}
\toprule
\multicolumn{3}{c|}{} & \multicolumn{3}{c|}{Baseline model} & \multicolumn{3}{c}{Text-prior model} \\
\cmidrule(lr){4-6} \cmidrule(lr){7-9}
$N$ & $J$ & $T$ & PAR & AAR$_1$ & AAR$_2$ & PAR & AAR$_1$ & AAR$_2$ \\
\midrule
800  & 10 & 1 & 0.980 (0.002) & 0.989 (0.001) & 0.991 (0.002) & 0.980 (0.002) & 0.989 (0.001) & 0.991 (0.001) \\
     &    & 2 & 0.947 (0.005) & 0.987 (0.002) & 0.958 (0.005) & 0.947 (0.004) & 0.987 (0.002) & 0.958 (0.004) \\
     & 20 & 1 & 0.985 (0.002) & 0.999 (0.000) & 0.986 (0.003) & 0.991 (0.002) & 0.999 (0.000) & 0.991 (0.002) \\
     &    & 2 & 0.979 (0.001) & 0.990 (0.002) & 0.989 (0.001) & 0.982 (0.003) & 0.990 (0.002) & 0.991 (0.003) \\
     & 30 & 1 & 0.903 (0.066) & 0.924 (0.067) & 0.941 (0.032) & 0.972 (0.006) & 1.000 (0.000) & 0.972 (0.006) \\
     &    & 2 & 0.940 (0.013) & 0.979 (0.017) & 0.943 (0.014) & 0.936 (0.015) & 0.998 (0.001) & 0.939 (0.014) \\
\addlinespace[0.3em]
1600 & 10 & 1 & 0.976 (0.002) & 0.985 (0.001) & 0.991 (0.001) & 0.976 (0.001) & 0.985 (0.001) & 0.991 (0.001) \\
     &    & 2 & 0.941 (0.005) & 0.987 (0.002) & 0.952 (0.005) & 0.943 (0.003) & 0.987 (0.002) & 0.955 (0.003) \\
     & 20 & 1 & 0.977 (0.010) & 0.997 (0.000) & 0.980 (0.010) & 0.992 (0.001) & 0.998 (0.001) & 0.994 (0.001) \\
     &    & 2 & 0.974 (0.006) & 0.993 (0.001) & 0.981 (0.006) & 0.980 (0.003) & 0.993 (0.001) & 0.986 (0.003) \\
     & 30 & 1 & 0.948 (0.030) & 0.973 (0.025) & 0.959 (0.021) & 0.940 (0.030) & 0.972 (0.027) & 0.951 (0.020) \\
     &    & 2 & 0.933 (0.017) & 0.993 (0.004) & 0.936 (0.018) & 0.936 (0.012) & 0.992 (0.005) & 0.939 (0.012) \\
\addlinespace[0.3em]
2400 & 10 & 1 & 0.978 (0.001) & 0.988 (0.001) & 0.990 (0.001) & 0.978 (0.001) & 0.988 (0.001) & 0.990 (0.001) \\
     &    & 2 & 0.957 (0.003) & 0.987 (0.001) & 0.969 (0.002) & 0.957 (0.003) & 0.987 (0.001) & 0.969 (0.002) \\
     & 20 & 1 & 0.989 (0.002) & 0.998 (0.000) & 0.990 (0.002) & 0.985 (0.004) & 0.998 (0.000) & 0.986 (0.004) \\
     &    & 2 & 0.976 (0.004) & 0.992 (0.001) & 0.984 (0.003) & 0.978 (0.002) & 0.992 (0.001) & 0.986 (0.001) \\
     & 30 & 1 & 0.931 (0.029) & 0.956 (0.029) & 0.948 (0.019) & 0.940 (0.025) & 0.969 (0.025) & 0.954 (0.015) \\
     &    & 2 & 0.912 (0.026) & 0.967 (0.027) & 0.924 (0.017) & 0.897 (0.032) & 0.944 (0.035) & 0.922 (0.019) \\
\bottomrule
\end{tabular}
\end{table}

Tables \ref{tab:gs_recovery} and \ref{tab:regression_recovery_baseline_textprior} summarize the estimation accuracy for item parameters ($g_{j,t}$, $s_{j,t}$) and model parameters ($\beta_0$, $\beta_Z$, and $\gamma_{01}$). Overall, estimation errors (RMSE and MAE) remained low across all conditions, with slightly higher errors under smaller sample sizes and shorter tests. Accuracy improved consistently as sample size ($N$) and number of items ($J$) increased, while bootstrap standard errors remained small, indicating stable estimation. Compared with the baseline model, the text-prior model generally achieved comparable or improved accuracy, particularly under more challenging conditions, where it gave lower estimation errors for both item parameters and regression parameters.

\begin{table}[!h]
\centering
\caption{Estimation accuracy of guessing and slipping parameters across time points ($T$) under varying sample sizes ($N$) and numbers of items ($J$), comparing the Baseline model and the Text-prior model. Values are reported as mean (bootstrap SE). The reported metrics include RMSE and MAE for guessing ($g$) and slipping ($s$) parameters.}
\label{tab:gs_recovery}
\footnotesize
\setlength{\tabcolsep}{3pt}
\begin{tabular}{ccc|cccc|cccc}
\toprule
\multicolumn{3}{c|}{} & \multicolumn{4}{c|}{Baseline model} & \multicolumn{4}{c}{Text-prior model} \\
\cmidrule(lr){4-7} \cmidrule(lr){8-11}
$N$ & $J$ & $T$ & $g$ RMSE & $g$ MAE & $s$ RMSE & $s$ MAE & $g$ RMSE & $g$ MAE & $s$ RMSE & $s$ MAE \\
\midrule
800  & 10 & 1 & 0.013 (0.001) & 0.011 (0.001) & 0.020 (0.003) & 0.016 (0.002) & 0.013 (0.001) & 0.011 (0.001) & 0.022 (0.003) & 0.017 (0.002) \\
     &    & 2 & 0.020 (0.002) & 0.016 (0.002) & 0.021 (0.001) & 0.017 (0.001) & 0.020 (0.002) & 0.016 (0.001) & 0.020 (0.001) & 0.017 (0.001) \\
     & 20 & 1 & 0.019 (0.002) & 0.015 (0.002) & 0.026 (0.003) & 0.019 (0.002) & 0.018 (0.002) & 0.014 (0.001) & 0.026 (0.003) & 0.019 (0.001) \\
     &    & 2 & 0.020 (0.002) & 0.016 (0.001) & 0.019 (0.001) & 0.015 (0.001) & 0.019 (0.002) & 0.015 (0.001) & 0.019 (0.001) & 0.015 (0.001) \\
     & 30 & 1 & 0.037 (0.006) & 0.030 (0.005) & 0.030 (0.002) & 0.021 (0.001) & 0.027 (0.003) & 0.022 (0.002) & 0.029 (0.002) & 0.022 (0.001) \\
     &    & 2 & 0.045 (0.006) & 0.031 (0.004) & 0.021 (0.003) & 0.016 (0.002) & 0.052 (0.004) & 0.033 (0.002) & 0.020 (0.001) & 0.016 (0.001) \\
\addlinespace[0.3em]
1600 & 10 & 1 & 0.012 (0.001) & 0.009 (0.001) & 0.017 (0.001) & 0.014 (0.001) & 0.012 (0.001) & 0.009 (0.001) & 0.017 (0.002) & 0.014 (0.001) \\
     &    & 2 & 0.016 (0.003) & 0.012 (0.002) & 0.015 (0.001) & 0.012 (0.001) & 0.014 (0.002) & 0.011 (0.001) & 0.015 (0.001) & 0.012 (0.001) \\
     & 20 & 1 & 0.019 (0.004) & 0.016 (0.003) & 0.021 (0.001) & 0.016 (0.001) & 0.014 (0.003) & 0.011 (0.002) & 0.020 (0.001) & 0.015 (0.001) \\
     &    & 2 & 0.021 (0.004) & 0.015 (0.002) & 0.015 (0.001) & 0.012 (0.000) & 0.016 (0.002) & 0.012 (0.001) & 0.015 (0.001) & 0.011 (0.001) \\
     & 30 & 1 & 0.021 (0.004) & 0.017 (0.003) & 0.021 (0.001) & 0.016 (0.001) & 0.028 (0.006) & 0.021 (0.004) & 0.021 (0.001) & 0.016 (0.001) \\
     &    & 2 & 0.047 (0.007) & 0.029 (0.004) & 0.014 (0.001) & 0.011 (0.001) & 0.043 (0.005) & 0.027 (0.003) & 0.014 (0.001) & 0.011 (0.001) \\
\addlinespace[0.3em]
2400 & 10 & 1 & 0.008 (0.001) & 0.006 (0.001) & 0.016 (0.002) & 0.013 (0.001) & 0.007 (0.001) & 0.006 (0.000) & 0.017 (0.002) & 0.013 (0.001) \\
     &    & 2 & 0.010 (0.001) & 0.008 (0.001) & 0.011 (0.001) & 0.010 (0.001) & 0.010 (0.001) & 0.008 (0.001) & 0.011 (0.001) & 0.010 (0.001) \\
     & 20 & 1 & 0.010 (0.001) & 0.008 (0.001) & 0.015 (0.001) & 0.012 (0.001) & 0.014 (0.003) & 0.011 (0.002) & 0.015 (0.001) & 0.012 (0.001) \\
     &    & 2 & 0.017 (0.002) & 0.012 (0.001) & 0.011 (0.000) & 0.009 (0.000) & 0.019 (0.003) & 0.013 (0.002) & 0.012 (0.001) & 0.009 (0.001) \\
     & 30 & 1 & 0.024 (0.004) & 0.018 (0.003) & 0.016 (0.001) & 0.012 (0.000) & 0.024 (0.003) & 0.018 (0.002) & 0.016 (0.001) & 0.012 (0.000) \\
     &    & 2 & 0.044 (0.005) & 0.026 (0.002) & 0.012 (0.001) & 0.009 (0.001) & 0.042 (0.004) & 0.026 (0.002) & 0.012 (0.001) & 0.009 (0.001) \\
\bottomrule
\end{tabular}
\end{table}

\begin{table}[!h]
\centering
\caption{Estimation accuracy of regression parameters under varying sample sizes ($N$) and numbers of items ($J$), comparing the Baseline model and the Text-prior model. Values are reported as mean (bootstrap SE) based on 1{,}000 bootstrap resamples. The reported metrics include root mean squared error (RMSE) and mean absolute error (mae) for $\beta_0$, $\beta_Z$, and $\gamma_{01}$. Lower values indicate better estimation accuracy.}
\label{tab:regression_recovery_baseline_textprior}
\footnotesize
\setlength{\tabcolsep}{4pt}
\begin{tabular}{cc|ccc|ccc}
\toprule
\multicolumn{2}{c|}{} & \multicolumn{3}{c|}{Baseline model} & \multicolumn{3}{c}{Text-prior model} \\
\cmidrule(lr){3-5} \cmidrule(lr){6-8}
$N$ & $J$ & $\beta_0$ & $\beta_Z$ & $\gamma_{01}$ & $\beta_0$ & $\beta_Z$ & $\gamma_{01}$ \\
\midrule
\multicolumn{8}{c}{\textit{RMSE (SE)}} \\
\midrule
800  & 10 & 0.104 (0.024) & 0.104 (0.006) & 0.102 (0.010) & 0.108 (0.025) & 0.105 (0.006) & 0.103 (0.010) \\
  & 20 & 0.094 (0.026) & 0.086 (0.006) & 0.095 (0.007) & 0.111 (0.031) & 0.084 (0.006) & 0.092 (0.004) \\
  & 30 & 0.275 (0.074) & 0.204 (0.049) & 0.141 (0.020) & 0.198 (0.039) & 0.119 (0.020) & 0.155 (0.015) \\
\addlinespace[0.3em]
1600 & 10 & 0.066 (0.011) & 0.064 (0.003) & 0.083 (0.007) & 0.068 (0.011) & 0.064 (0.003) & 0.079 (0.007) \\
 & 20 & 0.154 (0.037) & 0.104 (0.022) & 0.086 (0.006) & 0.102 (0.020) & 0.069 (0.011) & 0.074 (0.004) \\
 & 30 & 0.196 (0.042) & 0.101 (0.030) & 0.121 (0.016) & 0.245 (0.042) & 0.119 (0.040) & 0.118 (0.013) \\
\addlinespace[0.3em]
2400 & 10 & 0.044 (0.006) & 0.066 (0.003) & 0.062 (0.004) & 0.043 (0.007) & 0.064 (0.003) & 0.063 (0.005) \\
 & 20 & 0.077 (0.012) & 0.058 (0.003) & 0.066 (0.004) & 0.110 (0.019) & 0.070 (0.011) & 0.071 (0.007) \\
 & 30 & 0.177 (0.031) & 0.113 (0.035) & 0.132 (0.015) & 0.172 (0.030) & 0.124 (0.034) & 0.128 (0.013) \\
\midrule
\multicolumn{8}{c}{\textit{mae (SE)}} \\
\midrule
800  & 10 & 0.099 (0.021) & 0.087 (0.006) & 0.082 (0.007) & 0.105 (0.024) & 0.088 (0.006) & 0.084 (0.008) \\
  & 20 & 0.091 (0.025) & 0.068 (0.008) & 0.077 (0.006) & 0.106 (0.029) & 0.067 (0.007) & 0.074 (0.003) \\
  & 30 & 0.242 (0.060) & 0.164 (0.039) & 0.113 (0.016) & 0.176 (0.033) & 0.095 (0.017) & 0.123 (0.013) \\
\addlinespace[0.3em]
1600 & 10 & 0.058 (0.010) & 0.052 (0.002) & 0.066 (0.006) & 0.059 (0.011) & 0.051 (0.002) & 0.062 (0.006) \\
 & 20 & 0.143 (0.038) & 0.084 (0.017) & 0.068 (0.005) & 0.090 (0.018) & 0.057 (0.010) & 0.060 (0.003) \\
 & 30 & 0.174 (0.037) & 0.082 (0.023) & 0.085 (0.010) & 0.217 (0.040) & 0.097 (0.033) & 0.089 (0.012) \\
\addlinespace[0.3em]
2400 & 10 & 0.042 (0.006) & 0.053 (0.002) & 0.050 (0.004) & 0.041 (0.007) & 0.052 (0.003) & 0.050 (0.003) \\
 & 20 & 0.069 (0.011) & 0.047 (0.003) & 0.052 (0.003) & 0.097 (0.017) & 0.058 (0.009) & 0.059 (0.006) \\
 & 30 & 0.144 (0.028) & 0.091 (0.028) & 0.097 (0.012) & 0.143 (0.024) & 0.099 (0.027) & 0.098 (0.011) \\
\bottomrule
\end{tabular}
\end{table}

\section{Discussion}

We utilized NLP to construct a prior from item text information to inform the estimation of the $Q$-matrix. This helped improve model performance compared to models without a text-informed prior. Our proposal offers several advantages. First, it preserves the interpretability of the CDM framework. The NLP enters through a prior on $Q$-matrix complexity rather than through a black-box replacement of the measurement model. Second, it is based on the idea that text information may be useful but most likely imperfect in terms of determining the underlying attributes. Because the response likelihood remains central, misleading or uninformative text signals can in principle be overridden by the data. For example, in the present empirical application, the text-informed prior had limited practical effect on $Q$-matrix recovery, reflecting that the response data alone were sufficiently informative at this scale. The simulation study, conducted under more challenging conditions, provided stronger evidence for the prior's stabilising role. Third, the approach is especially promising in situations with dense item structures and settings where some rows are weakly identified from responses alone. In such cases, even modest prior information about plausible row complexity may improve stability and recovery. More broadly, this strategy illustrates how AI-based tools can support diagnostic modelling without sacrificing substantive interpretability. Previous studies \citep{delatorre2009cdm, sen2021sample} have emphasized the need for assessments of moderate length (e.g., at least 15 items) to ensure stable estimation. In contrast, our results suggest that the proposed model achieves reasonably good performance even with shorter tests, such as those with only 10 items, yielding acceptable levels of RMSE and bias. Additionally, the proposed model can be efficiently implemented using \texttt{nimble}\footnote{Code is available at: \url{https://osf.io/5rw8v/overview?view_only=5e263f9df22f45299f6e7528771a8510}.}, which facilitates relatively straightforward and computationally efficient estimation.

Several directions may extend the current modelling framework.
First, the choice of $\lambda$, which controls the strength of the text-derived prior, remains an open question. Misspecification of this parameter may lead to either over-reliance on noisy textual signals or underutilization of informative content. Developing a principled, data-driven approach for estimating $\lambda$ is therefore an important direction for future research. Second, in our simulation design, the $Q$-matrix was generated first, followed by the construction of text information $\tau$ conditional on $Q$. Although beyond the scope of this study, large language models (LLMs) could be used to generate simulated textual content for items and responses, from which $\tau$ can be extracted and incorporated into the proposed framework. This represents a promising direction for enhancing the realism of simulation studies. Third, the number of latent skills is typically determined based on the theoretical design of the Boost Reading program. While this study focuses on two skills, vocabulary and comprehension, the proposed framework can be extended to support data-driven selection of the number of skills. For example, clustering methods applied to text-based semantic representations (e.g., embedding-derived similarity measures) could be used to identify a more refined set of latent skills and corresponding item groupings, as explored in \citet{liu2026scalable} in a cross-sectional setting. 

In summary, the proposed methodology shows that NLP-derived item information can serve as a useful auxiliary source of evidence for $Q$-matrix estimation, pointing toward a broader class of hybrid models that combine advances in AI with the interpretability of cognitive diagnosis.

\section{Acknowledgments}

The authors declare no competing interests. This research was supported by the Economic and Physical Research Council, which funded the first author through a PhD studentship. The authors acknowledge the support of Boost Reading at Amplify for providing the dataset used in this analysis.

\section{Data availability}
Due to the commercial sensitivity of these data, our data sharing agreement with the company who provided the dataset requires that the raw data remain confidential and cannot be shared.

\vspace{\fill}\pagebreak




\bibliography{bibfile}
\vspace{\fill}\pagebreak

\section{Appendix}
\subsection{Data Preprocessing}

The empirical data were obtained from four game--grade combinations: Grade 2 \textit{Idiomatica}, Grade 3 \textit{Idiomatica}, Grade 2 \textit{Debate-a-ball}, and Grade 3 \textit{Debate-a-ball}. The total number of students in each dataset was as follows: 27,649 for Grade 2 \textit{Idiomatica}, 4,593 for Grade 3 \textit{Idiomatica}, 13,044 for Grade 2 \textit{Debate-a-ball}, and 18,220 for Grade 3 \textit{Debate-a-ball}. To construct a longitudinal sample, we identified students who appeared in both Grade 2 and Grade 3 datasets. This resulted in a final sample of 1,616 common students.

We conducted item selection separately for each game and grade based on student participation. For each game and grade, we computed (a) the number of students who attempted each item and (b) the proportion of participating students. Items were then ranked according to their frequency of appearance to identify those with the highest coverage. For each game--grade combination, we selected the top $k$ most frequently attempted items. The value of $k$ (e.g., 3, 5, 8, or 10) was determined by balancing the trade-off between including more items and retaining a larger sample size.

The selected items were:
\begin{itemize}
    \item Grade 2\textit{Idiomatica}: 1, 2, 3, 4, 5, 6, 7, 8, 9, 10
    \item Grade 3 \textit{Idiomatica}: 1, 2, 3, 4, 5, 6, 47, 48, 49, 51
    \item Grade 2 \textit{Debate-a-ball}: 37, 38, 39, 43, 44, 55, 81, 85, 86, 87
    \item Grade 3 \textit{Debate-a-ball}: 73, 74, 75, 79, 80, 81, 85, 86, 87, 100
\end{itemize}

We further restricted the dataset to students who completed all selected items across both games and both time points (Grade 2 and Grade 3). This ensured a balanced longitudinal structure for subsequent modelling.

\begin{figure}
    \centering
    \includegraphics[width=1\linewidth]{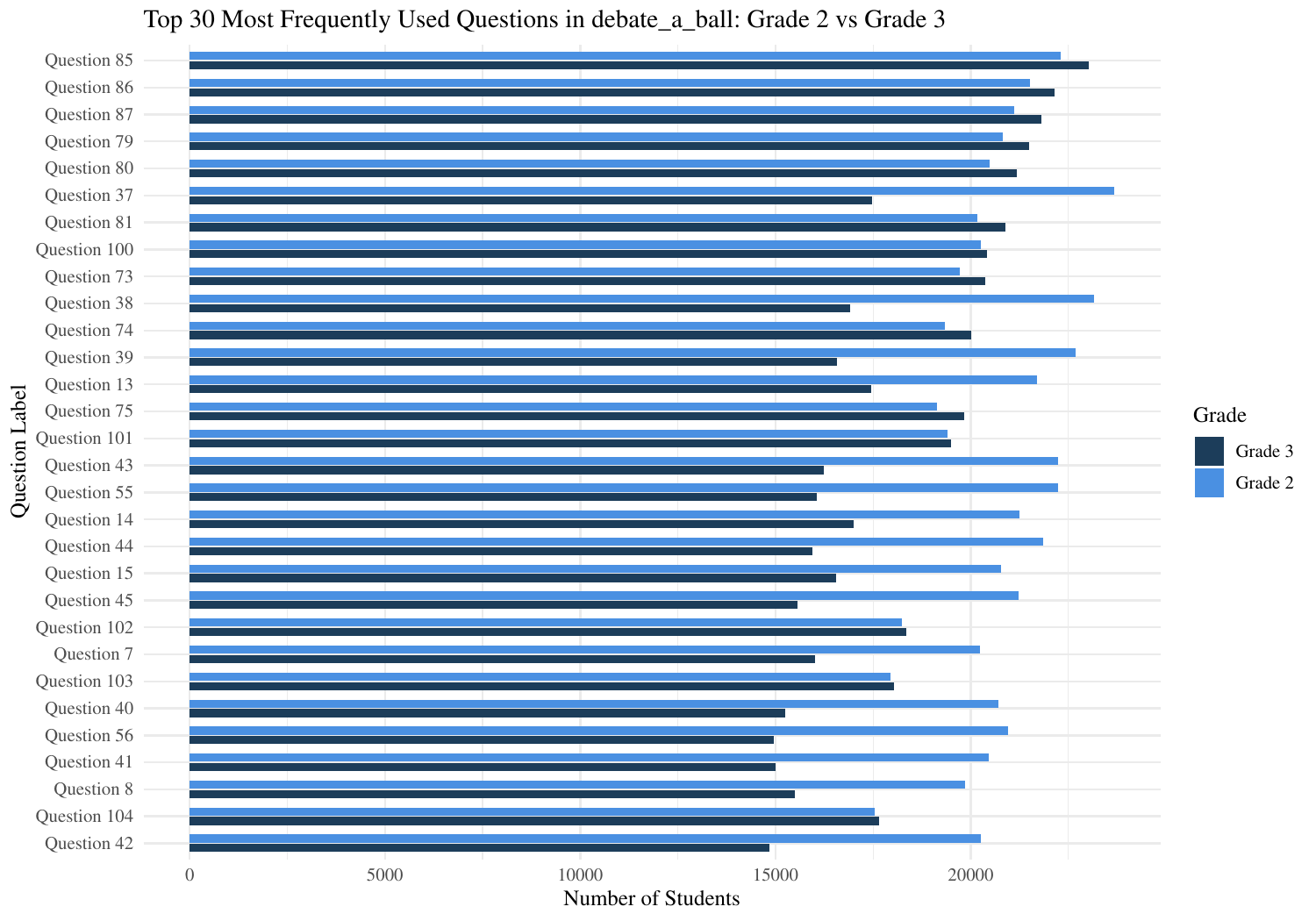}
    \caption{Top 30 most frequently attempted questions in Debate-a-ball, separately for Grade 2 and Grade 3. Bars show the number of students who attempted each question. Questions selected for the analysis are among the highest-frequency items.}
    \label{fig:debate_topgamesG12}
\end{figure}

\begin{figure}
    \centering
    \includegraphics[width=1\linewidth]{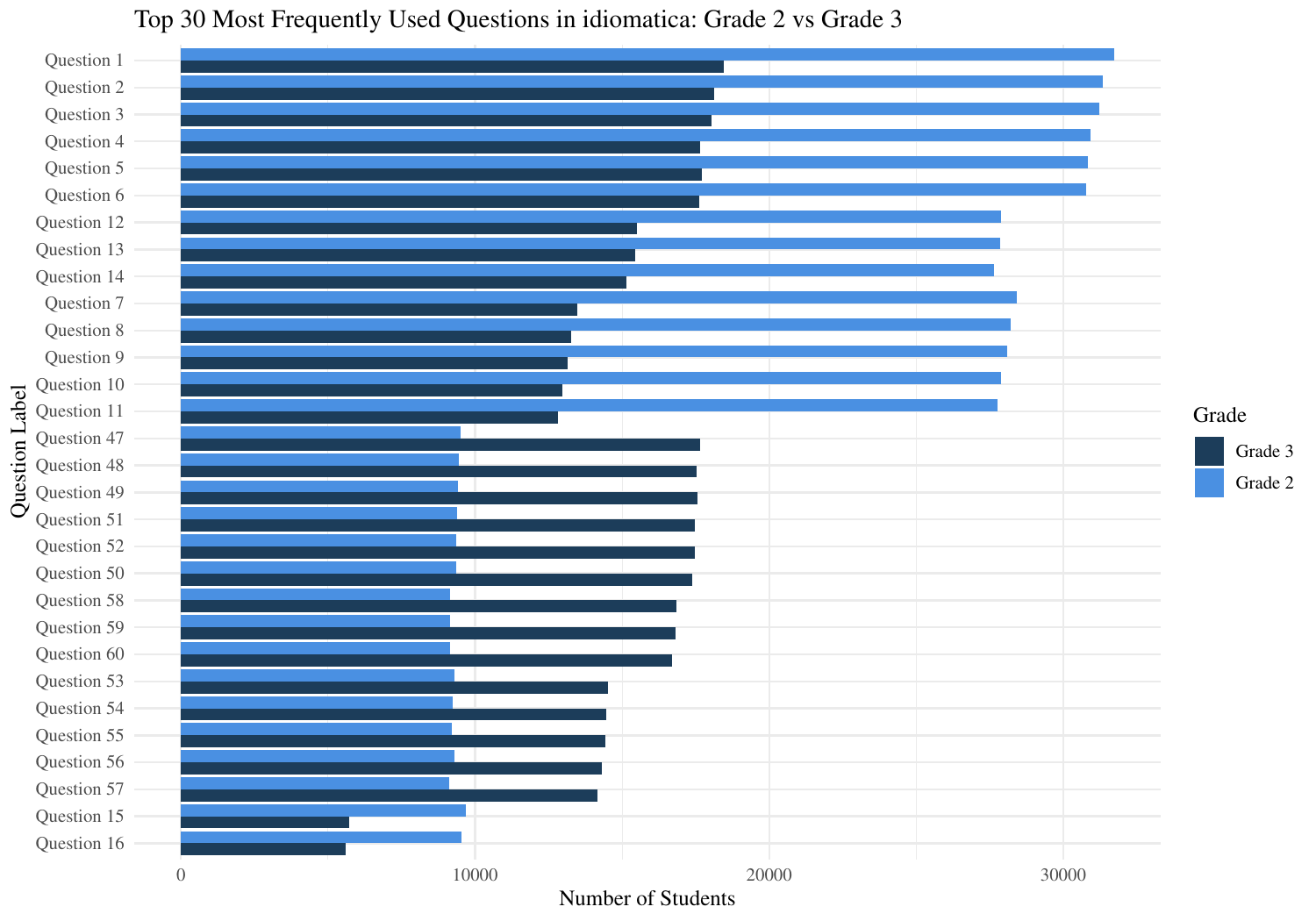}
    \caption{Top 30 most frequently attempted questions in Idiomatica, separately for Grade 2 and Grade 3. Bars show the number of students who attempted each question. Questions selected for the analysis are among the highest-frequency items.}
    \label{fig:idio_topgames}
\end{figure}

To identify the most informative subset of items, we conducted an exploratory data analysis (EDA) on question usage frequency. For each question, the number and proportion of students who attempted the item were computed. The top 30 most frequently used questions in each game are shown in Figures \ref{fig:debate_topgamesG12} and \ref{fig:idio_topgames}. Based on these distributions, a subset of high-frequency questions was selected to maximize sample size while maintaining sufficient coverage across levels.

The covariates were derived from students’ gameplay data across the two games. The number of attempts was computed as the average number of attempts per level for each student, and then averaged across the two games. The number of questions corrected (NLM) was defined as the total number of questions answered correctly across both games. Response time (RT) was calculated as the average response time per level for each student and subsequently averaged across the two games. These definitions and computation procedures are consistent with those used in our previous study \cite{ma2026dynamic}.

\subsection{Distribution of Text Information}
\begin{figure}
    \centering
    \includegraphics[width=.8\linewidth]{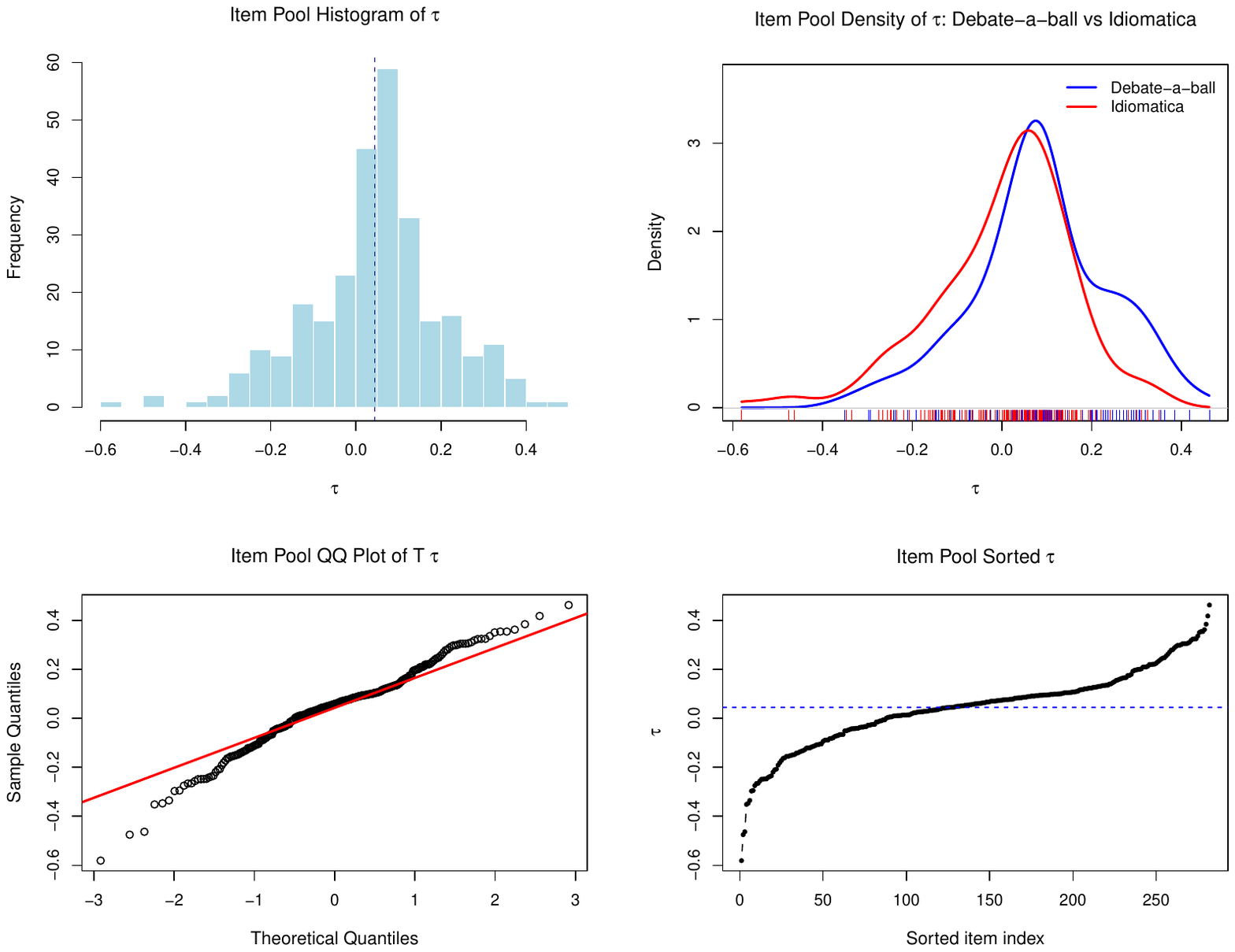}
    \caption{Distribution of $\tau$ values in the item pool.}
    \label{fig:Tdistri}
\end{figure}

\begin{figure}
    \centering
    \includegraphics[width=.8\linewidth]{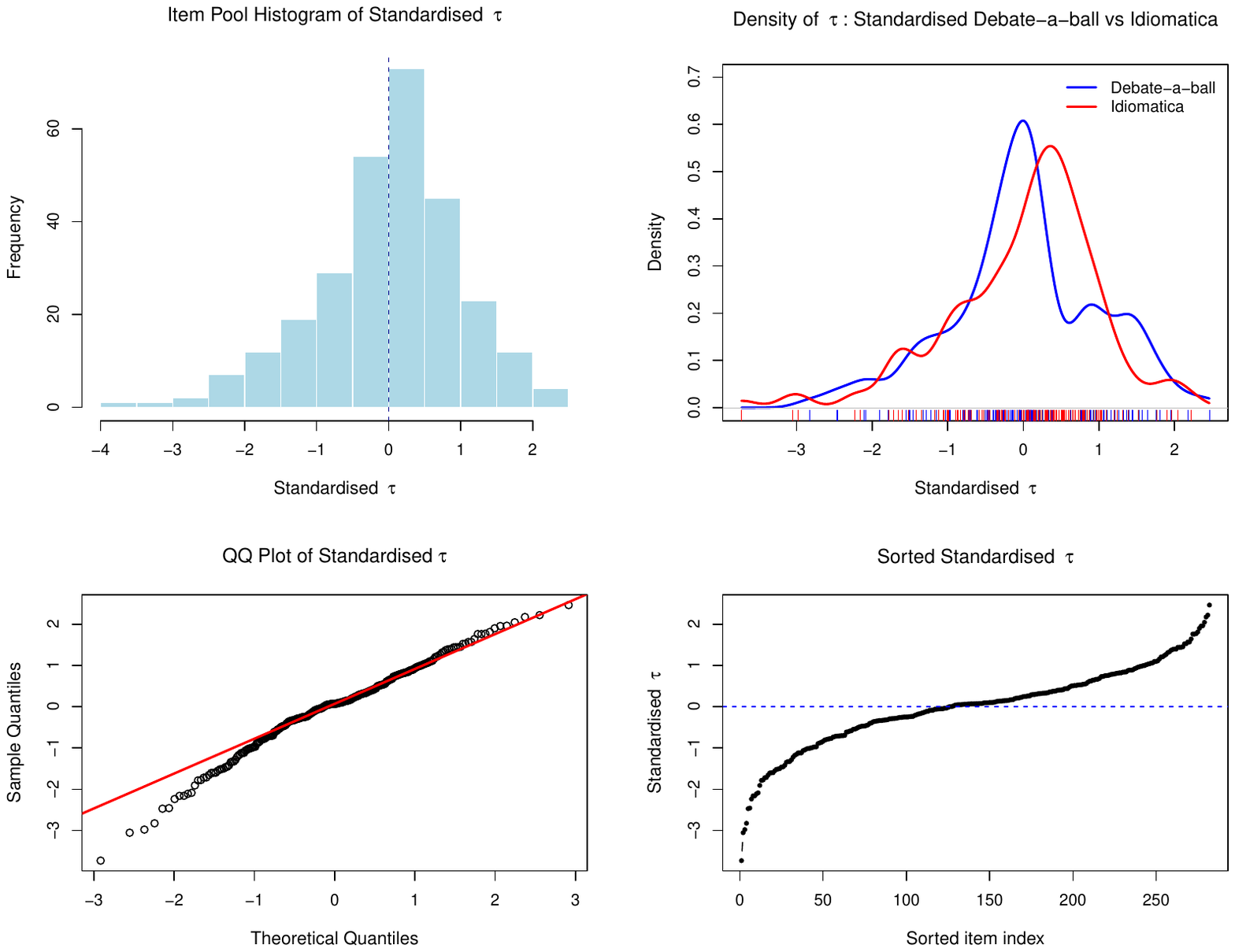}
    \caption{Distribution of standardised $\tau$ values in the item pool.}
    \label{fig:Tdistri_standard}
\end{figure}

The distribution of the item pool is shown in Figure \ref{fig:Tdistri}, and the standardized version is shown in Figure \ref{fig:Tdistri_standard}. Because both Grade 2 and Grade 3 were drawn from the same underlying item pool, the distribution of item-level $\tau$ is identical across grades. Differences between grades arise only from the item sampling process and subsequent student responses. The histogram and QQ plot indicate that $\tau$ is approximately normally distributed, with only mild deviations in the tails. The density plot further shows that debate items exhibit greater variability, whereas field items are more concentrated. The sorted values reveal a smooth distribution, providing no evidence of extreme outliers.

\subsection{True Q-matrix in simulation}
The shorter test forms were constructed as nested subsets of the full 30-item test. The true $Q$-matrices for the simulation with $J = 30$ and $K = 2$ are shown in Table~\ref{tab:qmatrix_j30_t1_t2}. The $Q$-matrices were held fixed across all simulation replicates.

\begin{table}[H]
\centering
\small
\caption{True $Q$-matrices for $J = 30$ at Time 1 and Time 2.}
\label{tab:qmatrix_j30_t1_t2}
\begin{tabular}{c|cc|cc}
\toprule
\multirow{2}{*}{Item} &
\multicolumn{2}{c|}{\textbf{Time 1}} &
\multicolumn{2}{c}{\textbf{Time 2}} \\
\cmidrule(lr){2-3} \cmidrule(lr){4-5}
& $A_1$ & $A_2$ & $A_1$ & $A_2$ \\
\midrule
1  & 1 & 0 & 1 & 0 \\
2  & 1 & 0 & 0 & 1 \\
3  & 0 & 1 & 1 & 0 \\
4  & 0 & 1 & 0 & 1 \\
5  & 1 & 0 & 1 & 1 \\
6  & 1 & 1 & 1 & 0 \\
7  & 0 & 1 & 1 & 0 \\
8  & 0 & 1 & 1 & 1 \\
9  & 1 & 0 & 1 & 1 \\
10 & 0 & 1 & 1 & 1 \\
11 & 1 & 0 & 1 & 1 \\
12 & 1 & 0 & 0 & 1 \\
13 & 1 & 0 & 1 & 1 \\
14 & 0 & 1 & 1 & 1 \\
15 & 1 & 1 & 1 & 1 \\
16 & 1 & 0 & 1 & 1 \\
17 & 1 & 1 & 1 & 1 \\
18 & 1 & 1 & 0 & 1 \\
19 & 1 & 1 & 1 & 1 \\
20 & 0 & 1 & 0 & 1 \\
21 & 0 & 1 & 1 & 1 \\
22 & 1 & 0 & 1 & 0 \\
23 & 0 & 1 & 0 & 1 \\
24 & 0 & 1 & 1 & 1 \\
25 & 0 & 1 & 1 & 1 \\
26 & 1 & 0 & 1 & 1 \\
27 & 0 & 1 & 1 & 1 \\
28 & 1 & 0 & 0 & 1 \\
29 & 1 & 1 & 1 & 0 \\
30 & 1 & 1 & 1 & 1 \\
\bottomrule
\end{tabular}
\end{table}



%



\end{document}